\documentclass[11pt]{article}

\pdfoutput=1

\usepackage{amsmath,amssymb}
\usepackage{epsfig}
\usepackage{slashed}
\usepackage{color}
\usepackage{graphicx}
\usepackage{subfigure}
\usepackage{fullpage}
\usepackage{tikzfeynman}
\usepackage{cite}

{\left\lbrace\begin{array}{@{}l@{}}}%
{\end{array}\right.}

\def\hc#1{{#1}^\dagger}
\def\ii{\mathit{i}}

\title{
\vspace{-2cm}
\vspace{3cm}
\bf \huge
Non Minimal Terms\\ in Composite Higgs Models and in QCD
\vspace{.5cm}}
\date{}
\author{
{\large Stefania De Curtis$^{a}$\footnote{stefania.decurtis@fi.infn.it},~ Michele Redi$^{a}$\footnote{michele.redi@fi.infn.it},~
Elena Vigiani$^{b,c}$\footnote{e.vigiani@studenti.unipi.it}}\\
[10mm] \normalsize\itshape $^a$ INFN, Sezione di Firenze, Via G. Sansone, 1; I-50019 Sesto Fiorentino, Italy
\\
\normalsize\itshape
$^b$ Universit\`a di Pisa, Dipartimento di Fisica, Largo B. Pontecorvo, 3; I-56127 Pisa, Italy\\
 \normalsize\itshape$^c$ INFN, Sezione di Pisa, Largo B. Pontecorvo, 3; I-56127 Pisa, Italy
}

\begin{document}
\maketitle
\begin{abstract}
\medskip
\noindent
We introduce a general parametrisation for theories where the Higgs is a Goldstone boson that encompasses all existing models.  
Our construction deviates from extra-dimensional or deconstructed theories through the inclusion of ``non-local'' interactions in theory space. 
These terms are necessary to reproduce the most general 4D effective lagrangian compatible with the symmetries.
After showing the relation of our formalism to the Callan-Coleman-Wess-Zumino effective lagrangian 
we apply our framework to $SO(5)/SO(4)$ composite Higgs models studying the implications for the Higgs potential, 
coupling of resonances and S-parameter. We also outline the relevance of non-minimal terms in effective descriptions of QCD, 
studying the electro-magnetic splitting of pions and the correlation between $L_9$ and $L_{10}$ in the chiral lagrangian.
\end{abstract}

\newpage
\tableofcontents

\section{Beyond Extra-Dimensions}

One of the most challenging areas of theoretical physics is the study of strongly coupled systems. 
The most practical tool is often the use of effective lagrangians that describe the interactions of light degrees
of freedom in an energy expansion that is strongly constrained by the underlying symmetries of the theory.
In particular for Goldstone bosons (GBs) arising from the spontaneous breaking of a global symmetry, 
the  Callan-Coleman-Wess-Zumino (CCWZ) construction \cite{CCWZ1,CCWZ2} provides a general parametrisation 
of the low energy dynamics that makes the symmetries of the theory manifest.

While the CCWZ construction describes the most general interactions compatible with the 
symmetries within a consistent effective field theory expansion valid up to a cut-off $\Lambda$, 
no dynamical information on the theory is retained. For example extra-dimensional theories
such as \cite{MCHM} can be described using the CCWZ formalism but properties such a locality in 5D are completely 
hidden from this point of view. Related to this fact, observables not constrained by the symmetries such as the Higgs potential, 
are not calculable. To make quantitative predictions one often needs to make extra dynamical assumptions on the size of
various operators. Moreover  the CCWZ approach appears useful only if very few degrees of freedom exist 
separated by a large gap from the rest of the dynamics, an assumption that can be violated in practice.

In this paper we provide a different parametrisation of theories with spontaneous symmetry breaking
that is well suited for composite models where the Higgs is a GB \cite{CHM4} or for the 
pions in QCD. Our starting point is Ref. \cite{4DCHM} where a minimal framework to describe the interactions of the 
lightest resonances with GBs was introduced (see also \cite{panico-wulzer} for a related construction). 
The dynamical assumption made in \cite{4DCHM} was the inclusion of only the nearest-neighbour interactions. 
In the limit of a large number of resonances, this construction becomes indistinguishable from an extra-dimension. 
Indeed various predictions, such as the Higgs mass \cite{higgsmass1,higgsmass2}, turn out to be similar to the 5D theories 
even with a minimal number of resonances.

We show that the most general effective lagrangian compatible with the symmetries is obtained 
by adding terms ``non-local'' in theory space to the lagrangian of Ref. \cite{4DCHM}. We will call these terms ``non-minimal'' since 
they allow to maximally deviate  from an extra-dimension\footnote{Non minimal terms also arise from the discretization of 5D theories with higher derivates with respect to the fifth coordinate \cite{serone}.}. These terms were also explored in \cite{wacker,thaler} where general moose
models were considered for QCD. In its most general form our lagrangian  is equivalent to the CCWZ one but our construction allows 
to control in a systematic way deviations from nearest-neighbour interactions. In particular the UV properties of quantities such as the 
Higgs potential are completely transparent being related to the notion of distance in the moose. For example we show that the 
Weinberg sum rules can be interpreted as constraints on the ``moose locality''.

We apply our formalism to the GB Higgs and to QCD. We derive in general interactions 
of heavy resonances, low energy lagrangian and two-point functions of currents and fermionic operators. 
We will show the impact of the non-minimal interactions on the GB potential and discuss the relation with other
effective lagrangians used to study these models at the LHC. For the GB Higgs we show that 
the tree level contribution to the $S-$parameter can be negative contrary to theories with nearest-neighbour interactions and compatibly 
with calculability of the potential. For QCD we show that the electro-magnetic splitting of pions and the KSRF relation \cite{bandokugo}
can be simultaneously reproduced with non minimal terms. We also show that, with the leading derivative interactions,
the parameters of the chiral lagrangian obtained integrating out the resonances satisfy $L_9 \simeq -L_{10}$ 
as suggested by experimental data.

The paper is organized as follows: in section \ref{model} we introduce the general two derivative effective lagrangian
with non minimal interactions. We consider vector and fermion resonances and show the equivalence
of our theory with the CCWZ construction. In section \ref{observables} we present formulas for various observables 
of interest in composite models and QCD. In section \ref{appGB} we discuss various implications of non-minimal terms
for the GB Higgs  and in section \ref{appQCD} for QCD. We conclude in section \ref{conclusions}. In the appendices 
we present the relevant formulas for composite fermions and a simplified model with non-minimal 
terms for spin-1 resonances.

\section{General 4D Moose Models}
\label{model}

We start by briefly reviewing the construction of Ref.\cite{4DCHM} to add resonances in a theory with a global 
symmetry $G$ spontaneously broken to a subgroup $H$.  For simplicity we focus on the cosets $SO(N)/SO(N-1)$, 
relevant for composite Higgs models and low energy QCD but our construction can be easily extended to general $G/H$ as in \cite{4DCHM}.

The spontaneous breaking $SO(N)/SO(N-1)$ can be parametrised by a unit $SO(N)$ vector\footnote{We adopt the normalization ${\rm Tr}[T^A T^B]=\delta^{AB}$ for the generators in the vectorial representation of $SO(N)$. We denote with $T^{a}$ the unbroken generators and with $T^{\hat a}$ the broken ones. The latter satisfy $\Phi^T_0 T^{\hat a} T^{\hat b} \Phi_0 =  \frac{1}{2} \delta^{\hat{a} \hat{b}}$.},
\begin{equation}
\Phi= U_0 \Phi_0
\end{equation}
where $\Phi_0=(0,\dots,0,1)^T$ and
\begin{equation}
U_0  = \exp{i \frac{\Pi(x)}{f}} , \quad \Pi=\sqrt{2}\ \pi^{\hat{a}}(x) T^{\hat{a}}
\end{equation}
is the GB matrix of the broken generators. The two derivative effective lagrangian for the GBs is just the kinetic term
\begin{equation}
\label{eqGBlagrangian}
{\cal L}= \frac{f^2}{2} \partial_\mu \Phi^T \partial_\mu \Phi \,
\end{equation}
where $f$ is the GB decay constant.

\subsection{Vectors}

We introduce composite spin-1 resonances to the GB lagrangian as gauge fields.
To this aim we add $K$ copies of non-linear $\sigma$-models, describing the spontaneous breaking $SO(N)_L^i\times SO(N)_R^i/SO(N)_{L+R}^i$. 
These are parametrised by orthogonal $SO(N)$ matrices transforming as
\begin{equation}
\Omega_i \to g_{L}^{i} \Omega_{i} (g_R^{i})^T \,, \quad i=1,\dots, K 
\end{equation}
while the unit vector $\Phi$ transforms as
\begin{equation}
\Phi \to g_L^{K+1} \Phi 
\end{equation}
and describes the spontaneous breaking $SO(N)_L^{K+1}/SO(N-1)$.
Next we gauge nearest-neighbour diagonal subgroups $SO(N)^{i}_R \equiv SO(N)_L^{i+1}$. 
The physical degrees of freedom are $K$ massive $SO(N)$ gauge fields interacting with the uneaten $SO(N)/SO(N-1)$ Goldstone bosons. 
Each gauge theory is associated with a site and the $\sigma$-models are described by the link fields between nearest-neighbour sites. 
This scenario  corresponds to the following lagrangian:
\begin{align}
\label{Lag_MIN}
\begin{split}
&{\cal L}= - \sum_{i=1}^{K}\frac{1}{4 g_i^2} \text{Tr}[A_{i}^{\mu\nu}A_{i \mu\nu}] + \sum_{i=1}^{K}\frac{f_{i}^{2}}{4} \text{Tr}[(D_{\mu}\Omega_{i})^{T}D^{\mu}\Omega_{i}] + \frac{f_{K+1}^{2}}{2}\left(D_{\mu}\Phi \right)^{T}\left(D^{\mu}\Phi \right)  
\end{split} \notag \\ \notag \\
&D_\mu \Omega_{i}= \partial_\mu \Omega_i - i A_\mu^{i-1} \Omega_i + i \Omega_i A_\mu^{i}\,, ~~~~~~~ D_\mu \Phi= \partial_\mu \Phi - i A_\mu^{K} \Phi \,.
\end{align}
The non dynamical fields $A_\mu^0$ are the sources of the global currents of the composite sector. Setting them to zero, the action has manifestly a global symmetry $SO(N)$ acting on the first site. The sources will be useful later on to compute correlation functions and to include elementary fields in the effective theory. 
For example in composite Higgs models SM gauge fields are introduced adding kinetic terms for the $SU(2)_L \times U(1)_Y$ sources. 
The global spontaneous symmetry breaking $SO(N)/SO(N-1)$ is induced by the last $\sigma$-model.
The physical GB matrix can be identified with
\begin{equation}
U= \prod_{i=1}^{K+1} \Omega_i  
\label{GBs}
\end{equation}
where $\Omega_{K+1}=U_0$.
The orthogonal scalar degrees of freedom are the longitudinal components of the massive spin-1 resonances.
To make manifest the particle content of the theory it is convenient to adopt the unitary gauge, where the GBs 
do not mix with the gauge resonances. Parametrizing: 
\begin{equation}
\Omega_i= \exp\left[i \frac{f \Pi(x)}{f_i^2}\right]\,, ~~~~~~~~ i=1, \dots, K+1 
\label{unitary_gauge}
\end{equation}
and comparing with (\ref{eqGBlagrangian}) one finds
\begin{equation}
\frac{1}{f^2}=\sum_{i=1}^{K+1} \frac{1}{f_i^2}  \,.
\end{equation}

\begin{figure}[t]
\begin{center}
\includegraphics[width=0.85\textwidth]{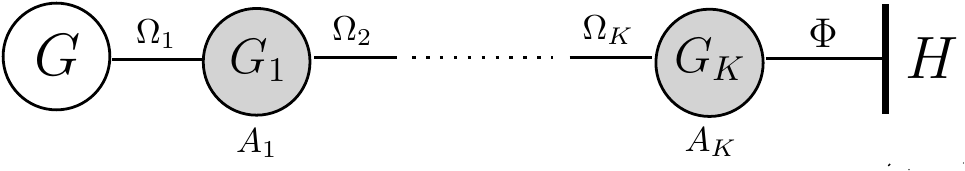}
\caption{\label{fig:minimal-moose} \small Moose model with nearest-neighbour interactions described by the lagrangian \eqref{Lag_MIN}. Shaded (white)
circles represent composite resonances (sources) while the GBs are the links between site $i$ and $i+1$. The spontaneous breaking $G/H$ is depicted
as a wall.}
\end{center}
\end{figure}

This construction naturally leads to the lagrangian with nearest-neighbour interactions and it is represented in Fig.~\ref{fig:minimal-moose}.
We will denote it as the minimal moose. This is similar to theories with one extra dimension and 
in fact the lagrangian  (\ref{Lag_MIN}) coincides with the discretization of a gauge theory in five dimensions. 
Not surprisingly the physical results resemble the ones found in 5D models
even for a small number of sites. 

This construction however does not  reproduce the most 
general lagrangian compatible with the symmetries  even at  two derivative order. 
Let us construct the products
\begin{equation}
\Omega_{i,j}\equiv \prod_{k=i+1}^j \Omega_k\,,~~~~~~~~~~~~~~~\quad i<j=1,\dots, K+1
\end{equation}
and define
\begin{equation}
\Omega_{j,i}\equiv (\Omega_{i,j})^T\,,~~~~~~~~~~~~~~~~~~ \Omega_{i-1, i}\equiv \Omega_i\,, ~~~~~~~~~~~~~~~~\Omega_{i ,i}\equiv {\cal I}
\end{equation}
The links $\Omega_{i,j}$ transform under the symmetries of sites $i$ and $j$ allowing them to directly communicate. 
At two derivate order we can write the following invariant terms:
\begin{equation}
\label{Lag_NM}
\begin{split}
{\cal L}&=\sum_{i,j=0}^{K}  \frac {f_{ij}^2}{8} \text{Tr}[D_\mu \Omega_{i,j} D^\mu \Omega_{j,i}] + \sum_{i=0}^{K} \frac{f_{i K+1}^{2}}{2} \Phi_0^T D_{\mu}\Omega_{K+1,i} D^{\mu}\Omega_{i,K+1}\Phi_0   \\
&-\sum_{i=1}^{K}\frac{1}{4 g_i^2} \text{Tr}[A_{i}^{\mu\nu}A_{i \mu\nu}]-\sum_{i=1}^K \frac 1 {2 \eta_{i}^2}  \Phi_0^T \Omega_{K+1,i}  A_i^{\mu\nu}A_{i \mu\nu}\Omega_{i,K+1} \Phi_0 \,.
\end{split}
\end{equation}
with
\begin{equation}
D_\mu \Omega_{i,j}= \partial_\mu \Omega_{i,j} - i A_\mu^i \Omega_{i,j} + i \Omega_{i,j} A_\mu^j 
\end{equation}
With this notation $f_{ij}$ is symmetric and $A_\mu^{K+1}=0$.   
The terms on the first line generate mass terms for the gauge fields allowing to reproduce the most general mass spectrum
compatible with the symmetries.  In the second line we wrote two possible invariants for the kinetic terms. 
Notice that the second one changes the normalisation of the kinetic terms of coset resonances and consequently their non abelian interactions.
For simplicity we do not include non minimal kinetic terms between site $i$ and $j$ that also modify the non-abelian interactions of vector resonances. 
In this language the minimal model (\ref{Lag_MIN}) corresponds to the choice  $f_{i-1i}=f_i$ and $\eta_i \rightarrow \infty$. From now on we will adopt the two indices notation also for the parameters of the minimal lagrangian.

The total number of parameters of the two derivative effective lagrangian (\ref{Lag_NM}) with $K$ $SO(N)$ resonances is:
\begin{equation}
\frac{K^{2}}{2}+\frac{7}{2}K+1
\end{equation}
We can also obtain a theory with a different number of $SO(N-1)$ and 
$SO(N)/SO(N-1)$ resonances for certain limits of parameters. In particular coset resonances can be decoupled by taking 
$f_{i K+1}$ to infinity.

Let us mention that the coset $SO(5)/SO(4)$, the minimal choice relevant for composite Higgs models, 
is special because the unbroken  subgroup is not simple. In fact $SO(4)\simeq SU(2)_L \times SU(2)_R$ 
so that there are two multiplets of resonances in the unbroken group transforming as $\bf{(3,1)\oplus(1,3)}$.
In this case another structure can be written down that distinguishes the two representations
\begin{equation}
\epsilon^{\alpha \beta \gamma \delta \rho}A_i^{A\mu\nu}A^B_{i\mu\nu} T^A_{\alpha \beta}T^B_{\gamma \delta} (\Omega_{i,K+1}\Phi_0)_\rho \,.
\label{LRbreaking}
\end{equation} 
This term breaks the symmetry that exchanges $SU(2)_L$ with $SU(2)_R$ with interesting phenomenological consequences \cite{Azatov:2013ura}.
Note that describing the resonances as gauge fields, no mass term can be written that breaks the $LR$ symmetry. 
For simplicity,  we will not include in our analysis non minimal terms that break $LR$ symmetry, 
actually considering the coset $O(N)/O(N-1)$.

The lagrangian (\ref{Lag_NM}) is the most general effective lagrangian up to two derivate order, compatible with the symmetries.
To show this property, it is useful to choose a gauge where the GBs appear in the fist link:
\begin{equation}
\label{CCWZgauge}
\Omega_{0,1}=U\,,~~~~~~~~~~~~~\Omega_{i-1,i}={\cal I}\,,~~~i=2\,,\dots\,K+1 \,.
\end{equation}
This naturally connects with the standard CCWZ parametrisation that makes manifest the invariance under shift of the GB lagrangian. 
Indeed in this gauge the only non derivative terms of the GBs appear in connection with the sources.
Let us  separate the $SO(N-1)$ resonances from the coset ones, indicating them as $\rho_{i}^{\mu}$ and $a_{i}^{\mu}$ respectively. 
In the gauge above the lagrangian reads:
\begin{equation}
\begin{split}
&{\cal L} = \sum_{j=1}^K \frac{f_{0 j}^2}{4} \text{Tr}\left[(e^\mu - \rho_j^\mu)^2 + (d^\mu - a_j^\mu)^2 \right] + \sum_{i,j=1}^K  \frac{f_{i j}^2}{8} \text{Tr} \left[(\rho_{i}^\mu - \rho_j^\mu)^2 + (a_{i}^\mu - a_j^\mu)^2 \right] \\
&+ \sum_{i=1}^{K} \frac{f_{i K+1}^2}{4} \text{Tr}[a_i^\mu a_{i\mu}] + \frac{f_{0 K+1}^2}{4} \text{Tr}[d^\mu d_\mu] \\
&-\sum_{i=1}^K \frac{1}{4 g_i^2} \text{Tr} \left[(\partial^\mu \rho_{i}^{\nu}-\partial^\nu \rho_i^\mu - i[\rho_{i}^{\mu},\rho_{i}^{\nu}] - i[a_{i}^{\mu},a_{i}^{\nu}])^2 \right] \\
&- \sum_{i=1}^K \frac{1}{4 \tilde{g}_{i}^{2}} \text{Tr} \left[(\partial^\mu a_i^\nu-\partial^\nu a_i^\mu - i[\rho_{i}^{\mu},a_{i}^{\nu}] - i[a_{i}^{\mu},\rho_{i}^{\nu}])^2 \right]
\label{LagCCWZ}
\end{split}
\end{equation}
where we have defined $1/\tilde{g}_{i}^{2}=1/g_i^2+1/\eta_i^2$. The symbols $e_\mu$ and $d_\mu$ are defined as usual
from the Maurer-Cartan form
\begin{equation}
i U^\dagger (\partial_\mu - i A^0_\mu) \ U = e^a_\mu T^a + d^{\hat a}_\mu T^{\hat a} = e_\mu + d_\mu \,.
\end{equation}
Diagonalizing the quadratic terms in $a_i$ and introducing kinetic terms for the elementary fields,
Eq.~(\ref{LagCCWZ}) coincides  with the lagrangian of Ref.\cite{GCHM} apart from the following subtlety. 
In Ref.\cite{GCHM} vector resonances transforming in the adjoint of the unbroken subgroup are introduced as gauge fields
while coset resonances are described as matter fields filling the fundamental representation of $SO(N-1)$ 
and their kinetic term are constructed with the covariant derivative $\nabla_\mu = \partial_\mu - i e_\mu$. 
In our model instead, all the resonances are described by gauge fields in the adjoint of $SO(N)$ 
and therefore extra-interactions are included from the non-abelian gauge interactions. 
Such interactions have important physical consequences as we will see.
We can recover the action of \cite{GCHM} integrating out the resonances of the unbroken group.  
The equation of motion of $\rho_i$ implies (to leading order for $p^2 \ll m_\rho^2$):
\begin{align}
\rho^\mu_i= e^\mu  \nonumber  \,.
\end{align}
From the non-abelian interactions of the gauge theory one reconstructs the covariant derivative of $a^i_\mu$.
However a spectrum with coset resonances significantly lighter than vector ones seems unlikely. 
For this reason we prefer to keep the $\rho$ resonances integrated in. 
In this way all the mathematics of non linear representation is encoded in the non abelian gauge interactions.

\subsection{Fermions}

Composite fermions can be treated in a very similar fashion. 
At each site $i$ we introduce Dirac fields $\Psi^i_r$ in a representation (in general reducible) of the local group.
The reps $r$ can vary from site to site. Fields at different sites can communicate through  the link fields $\Omega_{i,j}$. 

There are several ways to generalize the notion of minimal moose to fermions. In \cite{4DCHM} interactions with a chiral structure suggested by extra-dimension models were considered. For the third generation quarks the model corresponds to a moose with two Dirac fermions in the fundamental representation at each site.
Each fermionic source is associated to a composite fermion. The sources have a definite chirality and mix with fields with opposite chirality at the first site. The minimal moose is obtained by treating the fields at site $i$ as sources for the fields at site $i+1$. This induces a left-right chiral structure that in fact can be obtained by discretization of an extra-dimension. 

\begin{figure}[t]
 \begin{center}
\includegraphics[width=0.75\textwidth]{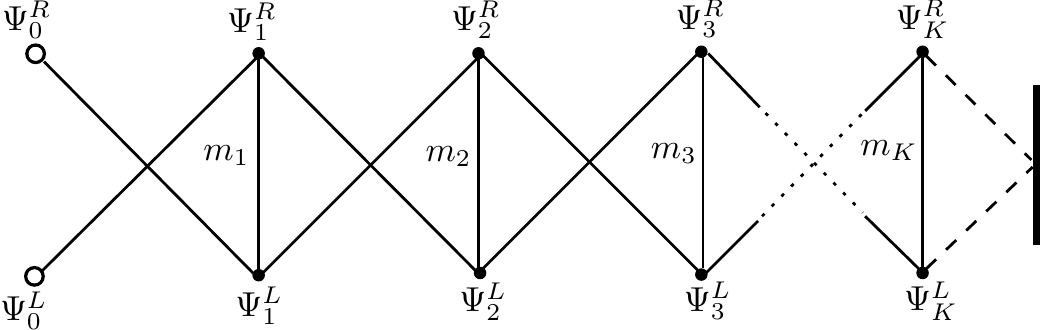}
 \caption{\label{fig:fermion-minimal-moose}\small Minimal fermionic moose with nearest-neighbour interactions.}
 \end{center}
\end{figure}

For simplicity here we will consider a single irreducible representation at each site. 
The minimal moose is characterized by only nearest-neighbour interactions and is depicted in Fig.~\ref{fig:fermion-minimal-moose}.
As done for the vector resonances, we want to exceed this minimal picture, introducing all the terms allowed by the symmetries.
Bilinears require fermions at different sites to belong to conjugate reps under the unbroken group. The number of invariants 
that can be written is then equal to the number of reps under the subgroup $SO(N-1)$.

Let us specialize to fermions in vector reps of $SO(N)$. Since there are 2 reps under $SO(N-1)$ (the vector representation and the singlet) the building blocks can be written as
\begin{equation}
{\cal O}_{ij}={\bar \Psi}_L^{i} \Omega_{i,j} \Psi_R^j\,,~~~~~~~~~~~{\cal O}'_{ij}={\bar \Psi}_L^i \Omega_{i,K+1}\Phi_0 \Phi^T_0 \Omega_{K+1,j} \Psi_R^j\,.
\end{equation}
To leading order, the fermionic lagrangian with non minimal interactions, in compact notation, is the following:
\begin{equation}
\label{L-ferm-NM}
{\cal L }=\sum_{i=1}^K \bar{\Psi}^i i \slashed{D}^{A_i} \Psi^i - \sum_{i,j=0}^K  \left[ M_{ij} {\cal O}_{ij}+  Y_{ij} {\cal O}'_{ij}  + h.c. \right]
\end{equation}
where $\Psi^0_L$ and $\Psi^0_R$ are the sources of fermionic operators in the vector representation 
of $SO(N)$. We do not include for simplicity symmetry breaking kinetic terms. 
These are however required to reproduce the most general effective lagrangian at two derivative order.

We note that non-minimal terms allow to interpolate between partial compositeness and
fermion masses \'a la technicolor. Partial compositeness corresponds to the hypothesis that
elementary fields acquire mass through the mixing to composite resonances. The term
\begin{equation}
\label{source-breaking}
{\cal O}'_{00}= \bar{\Psi}^0_L \Omega_{0,K+1}\Phi_0 \Phi_0^T \Omega_{K+1,0} \Psi^0_R 
\end{equation}
instead couples directly the fermionic sources to the VEV that breaks the global symmetry. 
Upon introducing elementary fermions this term generates fermion masses as in technicolor theories 
where the SM fermions couple to the fermion condensate that breaks the electro-weak symmetry.

\section{Physical Observables}
\label{observables}

In this section we provide the formulas needed to extract quantities of phenomenological relevance
from the effective lagrangian \eqref{Lag_NM}.  Our derivation generalizes the one in Refs.\cite{sonstephanov,becciolini}, 
in the context of moose models of QCD hadrons. Unless explicitly stated, we will neglect non minimal kinetic terms 
($\eta_i\to \infty$ or equivalently $\tilde{g_i}=g_i$).

\subsection{Resonance Couplings}

We can parametrise the physical GB fields with 
\begin{equation}
\begin{split}
&\Omega_{i} \equiv \Omega_{i-1,i}= \exp\left[i \omega_i  \Pi  \right] \,,~~~~~~~~ i=1, \dots, K+1 \\
&\sum_{i=1}^{K+1} \omega_i= \frac 1 f \,.
\end{split}
\label{smallomega}
\end{equation}
The coefficients $\omega_i$ depend on the gauge. To extract couplings to heavy spin-1 resonances it
is convenient to choose the unitary gauge where the physical GBs do not mix with the massive resonances. 
This implies $K$ equations for $K+1$ unknowns:
\begin{equation}
\sum_{j=0}^{K+1} \omega_{i j} f_{i j}^2 = 0\,,~~~~~~~~~~~~ i=1,\dots,K 
\end{equation}
where
\begin{equation}
\omega_{ij} =\sum_{k=i+1}^j \omega_k \,,~~~~~~~~~~~~~~~~~~\omega_{ji}=- \sum_{k=i+1}^j \omega_k = -\omega_{ij}
\end{equation}
The missing equation comes from the request  that the kinetic terms of the GBs are canonical
\begin{equation}
\frac 1 2\sum_{i,j=0}^{K+1} f_{ij}^2 \omega_{ij}^2= 1 \,
\end{equation}
that determines the physical decay constant $f$ in Eq.~(\ref{smallomega}).

In the spin-1 sector we define:
\begin{align}
\rho^\mu_k&= g_k \sum_{n=1}^K S_{kn} \tilde{\rho}^\mu_n \nonumber \\
a^\mu_k&= g_k \sum_{n=1}^K T_{kn} \tilde{a}^\mu_n
\label{eigenstates}
\end{align}
where $\tilde{\rho}^\mu_n$ and $\tilde{a}^\mu_n$ are the eigenstates with masses $m_{\rho_n}$ and $m_{a_n}$ respectively.
Masses and wave functions are determined by the eigenvalue problem
\begin{align}
\label{m_n}
2 S_{in} m_{\rho_n}^2&=\sum_{j=0}^K f_{i j}^2 g_i \left(g_i S_{in}-g_j S_{jn} \right)  \notag \\
2 T_{in} m_{a_n}^2&=\sum_{j=0}^{K} f_{i j}^2 g_i \left(g_i T_{in}-g_j T_{jn} \right)+ f_{i K+1}^2 g_i^2 T_{in}
\end{align}
with $i=1,\dots,K$ and  $g_0 = g_{K+1} = 0$.
Substituting (\ref{smallomega}) and (\ref{eigenstates}) 
into the lagrangian one can extract the couplings between GBs and vector resonances.
In particular we find
\begin{equation}
\label{grhopipi}
g_{\rho_n \pi\pi}= \sum_{i,j=0}^{K} \frac{f_{ij}^2}{2} \omega_{ij}^2 g_i S_{in} + \sum_{i=0}^K f_{i K+1}^2 \omega_{iK+1}^2 g_{i} S_{i n}
\end{equation}
where we have defined $g_{\rho_n \pi\pi}$ as the coefficient of the operator $f^{a \hat{b} \hat{c}}  \tilde{\rho}_n^{a \mu} \pi^{\hat b} \partial_\mu \pi^{\hat c}$, with $f^{a \hat{b} \hat{c}}$ structure constants of $SO(N)$ algebra. Similar manipulations can be performed for the fermions to determine their couplings to the GBs. 

\subsection{Form Factors}
\label{formfactors}

For many purposes it is useful to perform the path integral over the composite sector fields and write a 1PI effective action for the sources. This allows
to compute the GB effective potential generated by the coupling of elementary fields to the composite sector. 
In this way the strong dynamics is encoded into a set of form factors that describe correlation functions of operators of the composite sector. 
This is particularly simple at quadratic order in the fields and in a constant GB background. 
This technique was originally developed in the context of 5D theories \cite{MCHM} but it is completely general and can be applied in our non-minimal case. Here we closely follow Refs. \cite{MCHM} and \cite{4DCHM}, at which we refer for details.

Let us consider global symmetry currents.
For $SO(N)/SO(N-1)$ in a constant GB background and at quadratic order in the sources,
the effective lagrangian takes the following form\footnote{The formulas of this section and appendices \ref{B}, \ref{C} are properly defined in  euclidean space. The analytic continuation to  Minkowski space is understood.},
\begin{equation}
\label{Leff}
{\cal L}=\frac{1}{2}P_{t}^{\mu\nu}\Bigl[\Pi_{0}(p^{2})\text{Tr}[A^0_{\mu}A^0_{\nu}]+\Pi_{1}(p^{2}) \Phi^{T}A^0_{\mu}A^0_{\nu} \Phi\Bigr] ~,
\end{equation}
where $P_{t}^{\mu\nu}\equiv \eta^{\mu\nu}-\frac{p^{\mu}p^{\nu}}{p^{2}}$ and $\Phi = e^{i \Pi/f} \Phi_0$. 

The form factors $\Pi_{0}$ and $\Pi_{1}$  are related to the correlation functions of the currents associated to the unbroken and broken generators
\begin{align}
\Pi_{a}(p^{2}) (P_t)^{\mu \nu} &\equiv \langle J_a^\mu (p) J_a^\nu (-p) \rangle  \\
\Pi_{\hat a}(p^{2}) (P_t)^{\mu \nu} &\equiv \langle J_{\hat a}^\mu (p) J_{\hat a}^\nu (-p) \rangle  \,.
\end{align}
The relations between the form factors and the current correlators are the following
\begin{align}
\label{Pi0}
\Pi_{0}(p^{2})&=\Pi_{a}(p^{2}) \\
\label{Pi1}
\Pi_{1}(p^{2})&=2[\Pi_{\hat a}(p^{2})-\Pi_{a}(p^{2})] \,.
\end{align}
$\Pi_{a,{\hat a}}(p^2)$ can be extracted integrating out at tree level the composite resonances from the model lagrangian \eqref{Lag_NM} evaluated on a vanishing GB background.  
As long as the sources couple to the nearest resonance, one finds
\begin{equation}
\Pi_{a,\hat a}(p^2) = \frac{g_{1}^{2} f_{01}^{4}}{4}[(p^2 {\cal I}-{\cal M}_{\rho,a})^{-1}]_{11}+\frac{f_{01}^{2}}{2}
\label{gaugeformfactors}
\end{equation}
where ${\cal M}_\rho$, ${\cal M}_a$ are the squared mass matrices of vector and coset resonances that only differ through the symmetry breaking terms 
$f_{i K+1}$ on the diagonal, see Eq.~\eqref{m_n}. 

Using elementary properties of matrices one can write the following formula
\begin{eqnarray}
\label{Pi1-den}
&\Pi_1(p^2)& = ~\frac{g_{1}^{2}f_{01}^{4} }{2} ~ [(p^{2}{\cal I}-{\cal M}_\rho)^{-1} \cdot ({\cal M}_{a}-{\cal  M}_\rho)\cdot (p^{2}{\cal I}-{\cal M}_a)^{-1}]_{11} \nonumber \\
&=& \frac {\cal N}{\text{Det} [p^{2}{\cal I}-{\cal M}_\rho] \text{Det} [p^{2}{\cal I}-{\cal M}_a]}= \frac {\cal N}{\prod_{i=1}^K (p^2-m_{\rho_i}^2)(p^2-m_{a_i}^2)}
\end{eqnarray}
where $m_{\rho_i}$ and $m_{a_i}$ are the physical masses of vector and coset resonances (which are functions of $g_i$ and $f_i$).
The numerator is expressed in terms of the cofactor matrices ${\cal C}_\rho$ and ${\cal C}_a$ built from the minors of $(p^{2}{\cal I}-{\cal M}_\rho)^{-1}$ and $(p^{2}{\cal I}-{\cal M}_a)^{-1}$: 
\begin{equation}
{\cal N}=\frac{g_{1}^{2} f_{01}^{4}}{2} \ [{\cal C}_\rho \cdot ({\cal M}_a-{\cal M}_\rho) \cdot {\cal C}_a]_{11} \,.
\end{equation} 
For the minimal moose, the squared mass matrices have the typical nearest-neighbour form and \\ $[{\cal M}_a-{\cal M}_\rho]_{ij} = \frac{f_{K K+1}^2 g_K^2}{2} \delta_{iK} \delta_{jK}$, thus one finds
\begin{equation}
\label{Pi1-num}
{\cal N}= \frac{ f_{KK+1}^2}{4^K} \ \prod_{i=1}^K g_i^4 f_{i-1i}^4
\end{equation}
that is independent on momentum.  With non minimal interactions a momentum dependence appear in the numerator.
A minor generalisation allows to consider interactions between the sources and multiple resonances.
In this case one finds,
\begin{equation}
\Pi_{a,\hat{a}}(p^2) = \sum_{i,j=1}^K \left[ \frac{g_i g_j f_{0i}^2 f_{0j}^{2}}{4}[(p^{2}I-{\cal M}_{\rho, a})^{-1}]_{ij}+\frac{f_{0i}^2}2 \delta_{ij}\right]\,.
\end{equation}
Formula \eqref{Pi1-den} continues to hold with the numerator which is now momentum dependent.

To introduce elementary fields we simply need to set to zero the non-dynamical sources and add kinetic terms for
the remaining ones in Eq.~\eqref{Leff}. The same technique can be applied to the fermion sector, the main results 
are collected in appendix \ref{B}. 

\subsection{Goldstone Boson Lagrangian}
\label{GBlagrangian}

A  related computation is the low energy lagrangian for the GBs including higher derivative terms.
This is again obtained integrating out  the resonances but in a space-dependent GB background.
The effective lagrangian is then presented as an expansion in powers of momentum
\begin{equation}
{\cal L}=\frac{f^2}{4} \text{Tr}[d_\mu d^\mu]+\sum_i c_i O_i \,.
\end{equation}
We will focus on the following operators that are generated integrating out the resonances at tree level\footnote{For $SO(5)/SO(4)$ the complete ${\cal O}(p^4)$ lagrangian made of 11 operators can be found in \cite{continovector}. In that case the operators can be classified according to their parity with respect to the discrete symmetries of the theory. The formulae presented here hold in that case provided that the symmetry breaking term (\ref{LRbreaking}) is set to zero.},
\begin{align}
\label{operators}
O_1=& (\text{Tr} [d_\mu d^\mu])^2  \nonumber \\
O_2=& \text{Tr} [d_\mu d_\nu] \text{Tr} [d^\mu d^\nu] \nonumber \\
O_4^+=& \text{Tr} [f_{\mu\nu}^+ \ i [d^\mu, d^\nu]] \nonumber \\
O_5^+=& \text{Tr} [(f_{\mu\nu}^-)^2]
\end{align}
where $f_{\mu \nu}\equiv U^\dagger A^0_{\mu\nu} U=(f_{\mu \nu}^+)^a T^a+(f_{\mu \nu}^-)^{\hat a} T^{\hat a}\equiv f_{\mu \nu}^+ + f_{\mu \nu}^-$.

In order to obtain the coefficients $c_i$,  let us integrate out the heavy vector fields from the lagrangian \eqref{LagCCWZ}. We need to solve the equation of motions for the resonances and this can be 
 done order by order in a $p^2$ expansion. As customary in chiral perturbation theory, the vector fields are treated as terms ${\cal O}(p)$, so the EOM solution can be expanded as 
\begin{align}
\rho_{i\mu}=\rho^{(1)}_{i\mu}+\rho^{(3)}_{i\mu} \notag \\
a_{i\mu}=a^{(1)}_{i\mu}+a^{(3)}_{i\mu} 
\end{align}
The first terms are solution of the EOMs at zero momentum.
In principle to obtain the ${\cal O}(p^4)$ lagrangian, we would need the solution to the second order. 
However as in  5D theories \cite{pw5D} the ${\cal O}(p^4)$ tree level effective lagrangian is obtained by plugging the solutions at zero momentum in the kinetic terms. This is because ${\cal O}(p^4)$ terms arising from the product of ${\cal O}(p)$ and ${\cal O}(p^3)$ solutions in the mass terms automatically vanish by the EOMs. 
The solutions at zero momentum are
\begin{eqnarray}
\rho^{(1)}_{i \mu}&=& e_\mu \nonumber \\ 
a^{(1)}_{i \mu}&=& \alpha_i d_\mu , \quad i=1,\dots,K
\end{eqnarray}
where the coefficients $\alpha_i$ are obtained from the linearized EOMs at zero momentum
\begin{equation}
\label{alpha_i}
\sum_{j=0}^{K+1} f_{i j}^2 \left(\alpha_i - \alpha_j\right) = f_{0 i}^2 \,, \quad i=1, \dots, K \,.
\end{equation}
Plugging them into the kinetic terms we find the coefficients of the operators given in \eqref{operators}\footnote{Useful identities can be found in \cite{continovector}. In particular we used $e_{\mu\nu}=i [d_\mu,d_\nu]+f_{\mu\nu}^+$ and $f_{\mu\nu}^-=\nabla_{[\mu} d_{\nu]}$ with $\nabla_\mu=\partial_\mu-i e_\mu$ and $\text{Tr}\left(T^{\hat a}T^{\hat b}T^{\hat c}T^{\hat d}\right)=\frac{1}{4}\text{Tr}\left(T^{\hat a}T^{\hat b}\right)\text{Tr}\left(T^{\hat c}T^{\hat d}\right)+\frac{1}{4}\text{Tr}\left(T^{\hat a}T^{\hat d}\right)\text{Tr}\left(T^{\hat c}T^{\hat b}\right)$.}
\begin{eqnarray}
c_1&=& - c_2= -  \sum_{i=1}^K \frac{[1-\alpha_i^2]^2}{8 g_i^2} \nonumber \\
c_4^+&=&-2 c_5^+=  -\sum_{i=1}^K \frac{[1-\alpha_i^2]}{2 g_i^2} \,.
\label{so5coefficients}
\end{eqnarray}
The non-abelian gauge interaction are crucial to obtain these identities. 
A different result would be obtained for example in the formalism of Ref.~\cite{GCHM}.
We note that the inclusion of non minimal kinetic terms (\ref{Lag_NM}) 
for the resonances eliminates  correlations between the last two coefficients: 
\begin{align}
c_4^+ =& -\sum_{i=1}^K \frac{[1-\alpha_i^2]}{2 g_i^2} \nonumber \\
c_5^+ =& \sum_{i=1}^K \frac{1}{4} \left[\frac{1}{g_i^2}-\frac{\alpha_i^2}{\tilde{g}_i^2}\right]  \,.
\label{so5coefficientsnew}
\end{align}
As we will see in \ref{sec:L9L10} these relations have interesting implications for the ${\cal O}(p^4)$ 
QCD chiral lagrangian.

\section{Application to the Goldstone Boson Higgs}
\label{appGB}

We now discuss some physical consequences of the non-minimal terms in the context of composite Higgs models
where the Higgs is a GB.  We will then consider the standard pattern $SO(5)/SO(4)$ that produces 4 GBs 
with the quantum numbers of the Higgs doublet.

\subsection{The potential}
\label{sec:potential}

The Higgs potential arises entirely from the couplings that explicitly break the global symmetry of the theory. 
Minimally these are the SM Yukawa and gauge couplings. From the low energy point of view, 
the contributions to the potential are divergent but they can become finite due to the presence of resonances. 
This happens in 5D theories, and can be understood in terms of 5D locality. Since the Higgs 
potential corresponds to a non-local operator in 5D and all UV divergences are local in a local quantum field theory the
potential ought to be finite.  This property of 5D theories is also valid for their 4D avatars with nearest-neighbour interactions. 
In fact as discussed in \cite{4DCHM,GCHM,pomarol}  a single $G$ gauge field is sufficient for the convergence of the effective 
potential at 1-loop order. We now generalize those results to non-minimal interactions.

The gauge contribution to the 1-loop effective potential (neglecting hyper-charge for simplicity), is given by \cite{MCHM}:
\begin{equation}
\label{gauge pontential}
V(h)_{\rm gauge} = \frac{9}{4}\int \frac{d^4Q}{(2\pi)^4}\log \left[1 + \frac{1}{4}\frac{\Pi_1(Q^2)}{\Pi_0(Q^2)} s_h^2 \right]
\end{equation}
where $s_h\equiv \sin \frac{h}{f}$ and  $Q^2=-p^2$ is the euclidean squared momentum.
To account for the kinetic term of the elementary gauge bosons we have made the replacement in Eq.~(\ref{Pi0}),
\begin{equation}
\Pi_{0}(Q^{2})\to \frac{Q^2}{g_0^2} +\Pi_{0}(Q^{2})
\end{equation}
The convergence of the integral depends on the UV asymptotics of the form factors (IR divergences  are instead regulated by the finite SM masses). 
$\Pi_0(Q^2)\sim Q^2/g_0^2$ in the UV therefore finiteness of the potential requires that $\Pi_1(Q^2)$ 
goes to zero faster than $1/Q^2$. This implies two conditions:
\begin{equation}
\label{WSR}
\Pi_1(Q^2)\stackrel{Q^2\to \infty}{\longrightarrow} \ 0 \quad \text{(I)},~~~~~~~~~~ Q^2 \Pi_1(Q^2)\stackrel{Q^2\to \infty}{\longrightarrow} \ 0 \quad \text{(II)} \,.
\end{equation}
The first constraint eliminates the leading quadratic divergence from gauge loops (analogous to the SM one)  while the latter removes 
the residual logarithmic divergence. In QCD like theories the equations above can be translated into a relation
between the  masses and decay constants of the mesons of the theory. These are known as Weinberg sum rules.

Finiteness of the potential and the Weinberg sum rules can be translated into a  statement on the "locality" of the moose interactions,
i.e. the notion of distance between different sites. Let us begin considering nearest-neighbour interactions. 
From Eqs.~\eqref{Pi1-den} and \eqref{Pi1-num} we can immediately extract 
the leading order contribution to $\Pi_1(Q^2)$ in the UV
\begin{equation}
\label{Pi1_leading}
\Pi_1(Q^2) \sim \frac{f_{KK+1}^2 \prod_{i=1}^K  g_i^4 f_{i-1i}^4}{4^K Q^{4K}}\,. 
\end{equation}
It follows that the 1-loop potential is finite for $K\ge 1$.

It is illuminating to derive this result  diagrammatically.  For this purpose it is convenient to choose a gauge where the 
GBs  appear on the last link, that is
\begin{equation}
\label{gauge}
\Omega_{i-1,i}={\cal I} \, (i=1,\dots,K) \,, \quad \Omega_{K,K+1}=U \,.
\end{equation}
In a constant GB background the only term that contains the GBs is the mass term,
\begin{equation}
 \frac {g_K^2 f_{KK+1}^2} 2  (\Omega_{K,K+1}\Phi_0)^T A_{K\mu} A_K^\mu \Omega_{K,K+1}\Phi_0 \,.
\label{GBIR}
\end{equation}
The effective action can be written in terms of the two point functions of $A^\mu_1$.
Writing this as $\int 1/2  A_{0\mu}^A \Sigma_{AB}^{\mu\nu} A_{0\nu}^B $ we have
\begin{equation}
\Sigma_{AB}^{\mu\nu}= -\frac{g_1^2 f_{01}^4 }{4} \langle A^{\mu}_{1A} A^{\nu}_{1B}\rangle +  \frac {f_{01}^2}2  \eta^{\mu\nu} \delta_{AB}.
\label{sigmaab}
\end{equation}
The correlator $\Pi_1(Q^2)$, defined in \eqref{Leff}, can be extracted from the transverse part of $\langle A^{\mu}_{1A} A^{\nu}_{1B}\rangle$ proportional to the group structure $\Phi^T T^A T^B \Phi$ ($\Phi=\Omega_{K,K+1}\Phi_0$).
To determine the UV behavior of this correlator, we can work in the mass insertion approximation.
In the gauge (\ref{gauge}) the Feynman rules extracted from the lagrangian \eqref{Lag_MIN} are the following (we are here adopting the two indices notation):
\begin{center}
\begin{tikzpicture}[line width=1 pt, scale=1.65]
       \draw[vector] (0,0)--(0.5,0);
       \node at (0,-0.25) {$A_{i \mu}^A$};
       \node at (0.6,0) {$\times$};
       \draw[vector] (0.7,0)--(1.2,0);
       \node at (1,-0.25) {$A_{i \nu}^B$};
       \node at (4.8,0) {$\begin{Large}=\, \eta_{\mu\nu} [(-g_i^2 \frac{f_{i-1 i}^2+f_{i i+1}^2}{2}+g_K^2 \frac{f_{K K+1}^2}{2} \delta_{ik}) \delta^{AB} -g_K^2 f_{K K+1}^2 \delta_{ik} \Phi^T T^A T^B \Phi ] \end{Large}$};
       \draw[vector] (0,-1)--(0.5,-1);
       \node at (0,-1.25) {$A_{i \mu}^A$};
       \node at (0.6,-1) {$\times$};
       \draw[vector] (0.7,-1)--(1.2,-1);
       \node at (1,-1.25) {$A_{i+1 \nu}^B$};
       \node at (2.4,-1) {$\begin{Large}= \, g_i g_{i+1} \frac{f_{i i+1}^2}{2} \delta^{AB} \eta_{\mu\nu} \end{Large}$};    
\end{tikzpicture} \\
\end{center}
\begin{center}
\hspace{-8.6cm}\begin{tikzpicture}[line width=1 pt, scale=1.65]
    \draw[vector] (-6.5,0)--(-5.5,0);
    \node at (-6.5,0.18) {$\begin{small} \mu, A \end{small}$};
    \node at (-5.5,0.18) {$\begin{small} \nu, B \end{small}$};
    \node at (-6,-0.25) {$A_{i,T}$};
    \node at (-4.4,-0.05) {$\begin{huge} = \, \frac{1}{Q^{2}} (P_T)_{\mu\nu} \delta^{AB} \end{huge}$};
\end{tikzpicture} 
\end{center}
\begin{figure}[t]
 \begin{center}
\includegraphics[width=\textwidth]{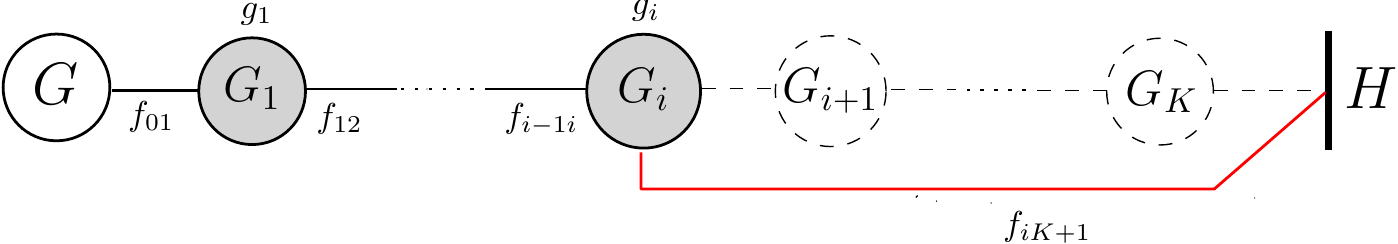}
 \caption{\label{fig:resonance-wall-link}\small Moose model with the non minimal link between the resonance $i$ and the symmetry breaking wall. The deep UV behaviour of the Higgs potential can be extracted from the minimal moose identified by the shortest path that includes the breaking. The dashed sites do not contribute to the UV leading order contribution on $\Pi_1(Q^2)$.}
 \end{center}
\end{figure}
All terms are diagonal in the global symmetry indices except for the one induced by Eq.~(\ref{GBIR}).
Obviously to generate contributions to the form factor $\Pi_1$ it is necessary to consider diagrams that include
the symmetry breaking vertex above. Since each propagator contributes with $1/Q^2$, the leading order 
contribution in the UV is generated by the diagram with the minimum number of insertions. 
This is obtained moving along the moose from the first site to the symmetry breaking wall and then going back to the first site:
\begin{center}
\begin{tikzpicture}[line width=1 pt, scale=1.65]
	\draw[vector] (2.6,0)--(3.4,0);
    \node at (2.9,0.25) {$A_1$};
    \node at (3.5,0) {$\times$};
    \draw[vector] (3.6,0)--(4.4,0);
    \node at (4.0,0.25) {$A_2$};
    \node at (4.7,0) {$\dots$};
    \draw[vector] (4.9,0)--(5.7,0);
    \node at (5.3,0.25) {$A_K$};
    \node at (5.8,0) {$\times$};
    \draw[vector] (5.9,0)--(6.7,0);
    \node at (6.3,0.25) {$A_K$};
    \node at (7.0,0) {$\dots$};
    \draw[vector] (7.2,0)--(8.0,0);
    \node at (7.6,0.25) {$A_2$};
    \node at (8.1,0) {$\times$};
    \draw[vector] (8.2,0)--(9,0);
    \node at (8.7,0.25) {$A_1$};
\end{tikzpicture}
\end{center}
Using the Feynman rules above, we obtain that the UV leading contribution to $\Pi_1(Q^2)$ is given by (\ref{Pi1_leading}).
The power of $Q^2$ is equal to the number of resonances  encountered along the path that starts from the source site, touches the breaking 
wall once and goes back to the sources. The number of resonances can then be defined as the length of the path along the moose. 
For the minimal moose $d_{AA}=2K$ so that $\Pi_1(Q^2) \sim 1/Q^{2d_{AA}}$. 

This result is immediately generalised to non-minimal terms that do not involve the sources. 
These terms, connecting non nearest-neighbour sites, reduce the distance between the sources 
and the symmetry breaking wall. As a consequence it is always possible to identify a shorter path 
inside the original moose that gives the leading contribution to $\Pi_1(Q^2)$.  Eq.~(\ref{Pi1_leading}) still holds for the minimal moose identified by the shortest path that goes from the source site to itself touching the breaking wall once. 
It is interesting to note that in order to study the UV convergence of the potential, it is always possible to reduce 
any moose to a nearest-neighbour one, as shown for example in Fig.~\ref{fig:resonance-wall-link}.

\begin{figure}[t]
 \begin{center}
\includegraphics[width=\textwidth]{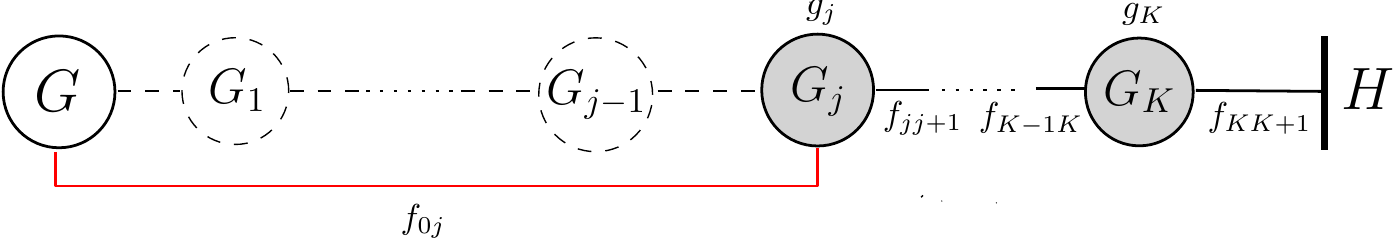}
 \caption{\label{fig:source-resonance-link}\small Moose model with the non minimal link between the source and the resonance $j$. The UV behaviour of the Higgs potential is determined by the shortest path involving $f_{0j}$.}
 \end{center}
\end{figure}

When the sources couple to more than one resonance, we have to be a little bit more careful. 
Consider for example a link between the source and the site $j$. Formula \eqref{sigmaab} is modified in
\begin{equation}
\Sigma_{AB}^{\mu\nu}= -\frac{g_1^2 f_{01}^4}{4} \langle A^{\mu}_{1A} A^{\nu}_{1B}\rangle -\frac{g_j^2  f_{0j}^4}{4} \langle A^{\mu}_{jA} A^{\nu}_{jB}\rangle -\frac{g_1 g_j f_{01}^2 f_{0j}^2}{2} \langle A^{\mu}_{1A} A^{\nu}_{jB}\rangle + \frac{f_{01}^2+f_{0j}^2}{2} \eta^{\mu\nu} \delta_{AB}    \,.
\end{equation}
It is not difficult to realize that the leading contribution comes from the diagram
\begin{center}
\begin{tikzpicture}[line width=1 pt, scale=1.65]
	\draw[vector] (2.6,0)--(3.4,0);
    \node at (2.9,0.25) {$A_j$};
    \node at (3.5,0) {$\times$};
    \draw[vector] (3.6,0)--(4.4,0);
    \node at (4.0,0.25) {$A_{j+1}$};
    \node at (4.7,0) {$\dots$};
    \draw[vector] (4.9,0)--(5.7,0);
    \node at (5.3,0.25) {$A_K$};
    \node at (5.8,0) {$\times$};
    \draw[vector] (5.9,0)--(6.7,0);
    \node at (6.3,0.25) {$A_K$};
    \node at (7.0,0) {$\dots$};
    \draw[vector] (7.2,0)--(8.0,0);
    \node at (7.6,0.25) {$A_{j+1}$};
    \node at (8.1,0) {$\times$};
    \draw[vector] (8.2,0)--(9,0);
    \node at (8.7,0.25) {$A_j$};
\end{tikzpicture}
\end{center}
corresponding to the shortest path along the moose that involves the breaking, with length $d_{AA}=2(K-j+1)$.
Also in this case formula \eqref{Pi1_leading} is valid with the shortest path defined by a linear moose (see Fig.~\ref{fig:source-resonance-link}).
Finally we note that the link $f_{0 K+1}$  trivially modifies $\Pi_{1}(Q^2) \rightarrow \Pi_{1}(Q^2)+f_{0 K+1}^2$, so that it goes to a constant at large  momenta. 
Again we can interpret this result in terms of the shortest path along the moose, that doesn't intersect any internal site and has vanishing length.

Keeping all the results together, we can conclude that 
\begin{equation}
\Pi_{1}(Q^{2}) \sim \frac{1}{Q^{2d_{AA}}}
\end{equation}
where the length $d_{AA}$ is defined as the number of resonances encountered along the shortest path that realizes the symmetry breaking.
The effective potential \eqref{gauge pontential} is always UV convergent except for the case $d_{AA}=0$, 
corresponding to the most non local interaction associated to $f_{0 K+1}$ that causes
a quadratic divergence in the potential. 

Let us now briefly consider models with incomplete $SO(5)$ multiplets. These can be recovered  by taking the limit $f_{i K+1} \rightarrow \infty$.
For nearest-neighbour interactions, using the explicit formulas (\ref{Pi1-den}), (\ref{Pi1-num}) and taking $f_{KK+1} \rightarrow \infty$, one finds:
\begin{equation}
\Pi_1(Q^2) \sim \frac{2 \prod_{i=1}^{K} g_i^4 f_{i-1 i}^4}{4^K g_K^2 Q^{4K-2}}
\end{equation}
i.e. the power is reduced by a factor $Q^2$ due to the decoupling of the last coset resonance. This result can be extended to non-minimal terms.
In this case we can conclude that if the shortest path includes the link $f_{i K+1}  \rightarrow \infty$, the UV behaviour of $\Pi_{1}(Q^{2})$ is modified in 
\begin{equation}
\Pi_{1}(Q^{2}) \sim \frac{1}{Q^{2d_{AA}-2}} \,.
\end{equation}
Thus a path of unit length can lead to a logarithmic divergence in the effective potential. An explicit example is considered in appendix \ref{C}.
Another possible source of  logarithmic divergences comes from non minimal kinetic terms. In this case the diagrammatic argument can still be applied, 
noting that they contribute with  $Q^2 \Phi^T T^A T^B \Phi$.

These results can be translated into the Weinberg sum rules \eqref{WSR}. 
The first  constraint requires $f_{0 K+1}=0$, i.e. the most non-local interaction in the moose 
should vanish. The second condition can be violated if incomplete $SO(N)$ multiplets are included or in the presence of non-minimal kinetic terms.
Let us also note that in some physical theories the second sum rule can be violated. This is the case of conformal technicolor \cite{orgogozo} where 
a non-integer power is obtained. This could be reproduced with our non local interactions but would require an infinite number of resonances.

Let us briefly discuss fermions. The effective action for composite fermions in the $\mathbf{5}$ rep of $SO(5)$ 
($\text{CHM}_{\mathbf{5}}$) in a constant GB background, takes the following form
\begin{eqnarray}
\label{CHM5-eff}
{\cal L}^{{\rm CHM_5}}_{{\rm eff}} &=&\ \!
 \bar{\Psi}_{0_L}^\alpha \slashed{p} \left( \delta^{\alpha \beta} \hat{\Pi}^{q_L}_0(p^2)
  + \Phi^\alpha \Phi^\beta \hat{\Pi}_1^{q_L}(p^2) \right)\! \Psi_{0_L}^\beta
  + \! \bar{\Psi}_{0_R}^\alpha \slashed{p} \left( \delta^{\alpha \beta}\hat{\Pi}_0^{u_R}(p^2)
  + \Phi^\alpha \Phi^\beta \hat\Pi_1^{u_R}(p^2) \right)\! \Psi_{0_R}^\beta \nonumber \\
 &+& \bar{\Psi}_{0_L}^\alpha \!\left( \delta^{\alpha \beta} \hat{M}_0^{u}(p^2)
   +\Phi^\alpha \Phi^\beta \hat{M}_1^{u}(p^2) \right)\! \Psi_{0_R}^\beta
 + h.c. 
\end{eqnarray}
We can follow exactly the same technique used for the gauge resonances to study the UV behaviour of the form factors appearing in the effective lagrangian \eqref{CHM5-eff}. For the minimal moose the effective lagrangian can be written in terms of two point functions of $\Psi_{1 L}$ and $\Psi_{1 R}$.
The form factors $\hat{\Pi}_1^{q_L}(p^2)$, $\hat{\Pi}_1^{u_R}(p^2)$, $\hat{M}_1^{u}(p^2)$ can be extracted  as the coefficient of the two point functions $\langle \Psi^1_L \bar{\Psi}^1_L \rangle$, $\langle \Psi^1_R \bar{\Psi}^1_R \rangle$ and $\langle \Psi^1_L \bar{\Psi}^1_R \rangle$ respectively, proportional to $\Phi^\alpha \Phi^\beta$.
From lagrangian \eqref{L-ferm-NM} specialized to nearest-neighbour interactions and with the GBs rotated in the last site, we obtain the Feynman rules ($\Phi=\Omega_{K,K+1} \Phi_0$):
\begin{center}
\begin{tikzpicture}[line width=1 pt, scale=1.65]
        \draw (0,0)--(0.5,0);
        \node at (0.1,-0.2) {$\bar{\Psi}_{iL}^\alpha$};
        \node at (.6,0) {$\times$};
        \draw (0.7,0)--(1.2,0);
        \node at (1.1,-0.2) {$\Psi_{iR}^\beta$};
        \node at (2.8,0) {$\begin{Large}{= -i m_i \delta^{\alpha \beta} - i Y_{KK} \delta_{iK} \Phi^\alpha \Phi^\beta}\end{Large}$};
        \draw (0,-1)--(0.5,-1);
        \node at (0.1,-1.2) {$\bar{\Psi}_{i-1 L}^\alpha$};
        \node at (.6,-1) {$\times$};
        \draw (0.7,-1)--(1.2,-1);
        \node at (1.1,-1.2) {$\Psi_{iR}^\beta$};
        \node at (2.1,-1) {$\begin{Large}{= -i M_{i-1 i} \delta^{\alpha \beta}}\end{Large}$};
\end{tikzpicture}
\end{center}
\begin{center}
\hspace{-3.5cm}
\begin{tikzpicture}[line width=1 pt, scale=1.65]
\draw[fermion] (0.8,0)--(-0.2,0);
        \node at (-0.3,0.1) {$\alpha$};
        \node at (0.9,0.1) {$\beta$};
        \node at (0.3,-0.2) {$\Psi_i$};
        \node at (1.6,-0.05) {$\begin{Large}{= \frac{i}{\slashed p} \, \delta_{\alpha \beta}}\end{Large}$};
\end{tikzpicture}
\end{center}
As we review in appendix \ref{B}, the convergence of the potential implies the following constraints
\begin{equation}
\begin{split}
&Q^4 \hat{\Pi}_1^{q_L}(Q^2)\stackrel{Q^2\to \infty}{\longrightarrow} \ 0 \quad , \quad Q^4 \hat{\Pi}_1^{u_R}(Q^2)\stackrel{Q^2\to \infty}{\longrightarrow} \ 0\\
&Q^2 |\hat{M}_1^u(Q^2)|^2 \stackrel{Q^2\to \infty}{\longrightarrow} \ 0 \,.
\end{split}
\label{weinbergsumrules}
\end{equation}
with $Q^2=-p^2$.
Analogously to the spin-1 sector, to obtain contribution to the form factors above, it is necessary to consider a diagram that involves the symmetry breaking terms on the last site. The UV leading order contribution to the form factors is  given as before by the shortest path along the moose that connects the sources, touching at least once the breaking wall. It is easy to obtain 
\begin{eqnarray}
\hat{\Pi}_1^{q_L}(Q^2) &\sim \frac {1} {Q^{d_{LL}+1}} \nonumber \\
\hat{\Pi}_1^{u_R}(Q^2) &\sim \frac {1} {Q^{d_{RR}+1}} \nonumber \\
\hat{M}_1^{u}(Q^2) & \sim \frac {1} {Q^{d_{LR}}}
\end{eqnarray}
where we introduced the fermionic distances  as the number of chiral fermions encountered
along the shortest path. Note that this distance  depends on the chirality.
Finiteness of the potential then translates into the following requirements on the distances in fermionic sector
\begin{equation}
d_{LL,RR} > 3\,~~~~~~~~~d_{LR} > 1 \,,
\end{equation}
meaning that we need two Dirac composite fields (four chiral components) to make the potential finite.

\subsection{Simplified Model}

To elucidate the role of non-minimal terms we now consider a simplified model with just a single multiplet of resonances, fermionic and bosonic.
This model essentially captures all the relevant features that could be accessible at the LHC.

\begin{figure}[t]
 \begin{center}
\includegraphics[width=0.3\textwidth]{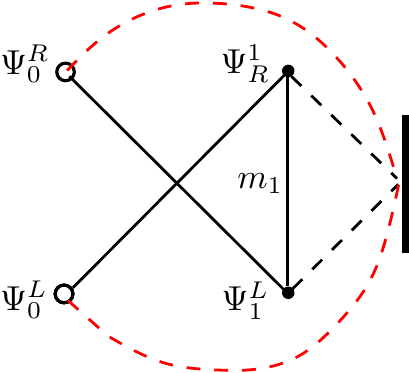}
 \caption{\label{fig:simplified_model}\small Schematic picture of the fermionic interactions in lagrangian \eqref{CHM5}, where red lines represent non minimal links.}
 \end{center}
\end{figure}
 The composite degrees of freedom are a Dirac fermion 
in the $\mathbf{5}$ rep and a complete $SO(5)$ gauge multiplet. For the bosonic sector the lagrangian is
\begin{equation}
{\cal L}_{gauge}=\frac{f_{01}^2}{4}{\rm Tr} \left[(D_{\mu}\Omega_{0,1})^T (D^{\mu}\Omega_{0,1})\right]+  \frac {f_{12}^2} 2 \left(D_{\mu}\Omega_{1,2} \Phi_0 \right)^T\left(D_{\mu}\Omega_{1,2} \Phi_0 \right) -\frac 1 {4 g_1^2}
{\rm Tr}\left[A_{1\mu\nu} A_1^{\mu\nu}\right]
\label{gaugeminimal}
\end{equation}
while the fermionic lagrangian reads (see  Fig.~\ref{fig:simplified_model}):
\begin{equation}
\label{CHM5}
\begin{split}
{\cal L}^{\rm CHM_5} &= 
-M_{01} \bar{\Psi}^0_{L} \Omega_{0,1} \Psi^1_{R} - M_{10}^* \bar{\Psi}^0_{R} \Omega_{0,1} \Psi^1_{L} +  h.c.  \\
& + \bar{\Psi}_1 (i \slashed{D}^{A_1} -m_1) \Psi_1 - Y_{11} \bar{\Psi}^1_{L} \Omega_{1,2} \Phi_0 \Phi_0^T \Omega_{2,1} \Psi^1_{R} + h.c. \\
& - Y_{01} \bar{\Psi}^0_{L} \Omega_{0,2} \Phi_0 \Phi_0^T \Omega_{2,1} \Psi^1_{R} - Y_{10}^*  \bar{\Psi}^0_{R} \Omega_{0,2} \Phi_0 \Phi_0^T \Omega_{2,1} \Psi^1_{L} + h.c. \,.
\end{split}
\end{equation}
In the last line we have written non-minimal terms that connect directly the elementary fields with the symmetry breaking. 
We have not included, for fermions and bosons, the direct coupling between the sources that would violate the 
hypothesis of partial compositeness. 
For $Y_{01}=Y_{10}=0$   the fermionic distances are
\begin{equation}
d_{LL}=d_{RR}=3\,~~~~~~~~~~~~d_{LR}=2 \,.
\end{equation}
It follows from the discussion above that the potential has a logarithmic divergence due to
the LL and RR contributions. On the contrary, the contribution to the potential proportional to the top Yukawa arising from the LR contribution, remains finite.
With non-minimal terms the fermionic distances $d_{LL}$ and $d_{RR}$ are reduced by two units
leading to quadratically divergent contributions to the potential. On the other hand the LR distance remains unchanged so that 
the contribution to the potential controlled by the top Yukawa coupling is still finite (this contribution becomes
quadratically divergent only in the presence of the interaction $Y_{00}$ in Eq.~(\ref{L-ferm-NM})
connecting left-handed and right-handed sources).  Note that in this case light fermionic partners are 
not sufficient for naturalness since the potential is not saturated by the lighter resonances.

Let us now discuss the limit where the fermionic resonances are lighter than the spin-1 resonances.
This configuration is often motivated by the naturalness of the theory even though the presence
of non minimal terms weakens the connection between light top partners and naturalness of the electro-weak scale. 
In this situation we can integrate out the spin-1 resonances. This is conveniently performed in the CCWZ gauge
\eqref{CCWZgauge}. To leading order,
\begin{equation}
\label{eom}
\rho_{1\mu} = e_\mu \,,~~~~~~~~~~~~~~a_{1\mu} = \frac{f_{01}^2}{f_{01}^2+f_{12}^2} d_\mu \,.
\end{equation}
Substituting into the lagrangian (\ref{gaugeminimal}) we obtain
\begin{equation}
\begin{split}
{\cal L} = \frac{1}{4}\frac{f_{01}^2 f_{12}^2}{f_{01}^2 + f_{12}^2} {\rm Tr}[d_\mu d^\mu]
\end{split}
\end{equation}
so that the physical decay constant of the GBs is $f^2=(f_{01}^2 f_{12}^2)/(f_{01}^2+f_{12}^2)$.
For the fermions we get
\begin{align}
\begin{split}
{\cal L}&= -M_{01} (\bar{\Psi}_{0L} U)_\alpha Q_{1R}^\alpha - (M_{01}+Y_{01}) (\bar{\Psi}_{0L} U)_{5} S_{1R} + \text{h.c.}  \\ 
&- M_{10}^* (\bar{\Psi}_{0R} U)_\alpha Q_{1L}^\alpha - (M_{10}^*+Y_{10}^*) (\bar{\Psi}_{0R} U)_{5} S_{1L} + \text{h.c.}  \\
& + \bar{Q}_1 \left(i \gamma^\mu \partial_{\mu} - \gamma^\mu e_\mu -m_1 \right)Q_1 + \bar{S}_1 \left(i \gamma^\mu \partial_{\mu}-(m_1+Y_{11})\right) S_1 \\
&+ \frac{i}{\sqrt{2}} \frac{f_{01}^2}{f_{01}^2+f_{12}^2} \bar{Q}_{1}^\alpha \gamma^\mu d_\mu^\alpha S_{1} + \text{h.c.} 
\end{split} 
\end{align}
where we distinguish between fourplet $Q_1^\alpha$,  $\alpha=1, \dots, 4$, and singlet $S_1$.
This lagrangian is equivalent to the one considered in \cite{grojean-matsedonskyi-panico} where the most general lagrangian of 
singlet and fourplet were studied, see also \cite{higgshunter}. Note that in our formalism the last term, 
that induces an interaction between the fourplet and the singlet, mediated by the $d_\mu$ symbol, 
originates from the covariant derivatives with the coset resonances. The coefficient 
\begin{equation}
\label{c}
c=\frac{1}{\sqrt{2}} \frac{f_{01}^2}{f_{01}^2+f_{12}^2} = \frac 1 {\sqrt 2} \frac {m_\rho^2}{m_{a_1}^2}
\end{equation}
is tunable but is always positive in our setup. 
In the minimal moose it is smaller than $1/\sqrt{2}$, but can be made larger
adding the non-local term that connects the sources,
\begin{equation}
\frac{f_{02}^2}{2} \left(D_{\mu} \Omega_{0,2} \Phi_0 \right)^T \left(D_{\mu} \Omega_{0,2} \Phi_0 \right)  \,.
\end{equation}
In this case in fact the coset resonances can be made lighter than the vector ones so that $c>1/\sqrt{2}$.

The typical size of the coefficient \eqref{c} agrees with the partial UV completion criterion advocated in Ref. \cite{continovector}.
Various limits considered in the literature can be recovered. For $f_{12}\rightarrow \infty$ the coset resonances are decoupled and $c=0$.
This is equivalent to the two site model in \cite{panico-wulzer}. For $f_{01}\rightarrow \infty$ all the vector resonances acquire infinite mass 
and one finds $c=1/\sqrt{2}$.  This corresponds to the model of Ref. \cite{furlan} also considered in \cite{grojean-matsedonskyi-panico}.

\subsection{$S$-parameter}

A severe constraint on theories where strong dynamics breaks the electro-weak symmetry arises from electroweak precision tests, 
in particular the $S$-parameter. In this section we will focus on the tree-level contribution arising from spin-1 resonances. The NDA estimate is given by
\begin{equation}
\Delta S \sim \frac {4 \pi v^2}{m_\rho^2} 
\label{NDA S}
\end{equation}
where $m_\rho$ is the scale of vector resonances. In extra-dimensional theories one can prove that the tree level contribution to $S$ is always positive \cite{S-rattazzi} and of the order above. Recently it was pointed out that sizable negative contributions can originate from fermions \cite{grojean-matsedonskyi-panico,Azatov:2013ura}.

The general expression of $\Delta S$ at tree level can be conveniently extracted from the
two-point functions of the currents of the composite sector \cite{MCHM},
\begin{equation}
\Delta S= 4\pi \frac{v^{2}}{f^{2}} \frac{d \ \Pi_{1}}{d p^2} \Big|_{p^2=0} = 4\pi v^{2} \frac{d \ \text{log} \Pi_{1}}{d p^2} \Big|_{p^2=0} \,.
\end{equation}
where we used the fact that $\Pi_1(0)=f^2$.
From Eq.~(\ref{Pi1-den}) it follows
\begin{equation}
\label{S_4DCHMN}
\Delta S =4\pi v^{2} \sum_{i=1}^{K} \left(\frac{1}{m_{\rho_i}^{2}}+\frac{1}{m_{a_i}^{2}}\right)   + 4\pi v^{2} \frac{d}{d p^2}  \text{log}{\ \cal N} \Big|_{p^2=0}  \,.
\end{equation}
With nearest-neighbour interactions ${\cal N}$ is momentum independent. In this case the second term is zero so that  $\Delta S$
is always positive and at least as big as (\ref{NDA S}), in agreement with the results in extra-dimensional theories.
Note that its expression only depends on the masses  of the resonances. This is a special feature of the GB Higgs 
that would not apply to a generic composite  state (see for example \cite{gersdorff}). 

Let us now turn to non minimal interactions where ${\cal N}$ is momentum dependent 
and interesting effects can be obtained. Indeed introducing the most non-local term one finds,
\begin{equation}
\Delta S= 4\pi v^{2} \sum_{i=1}^{N} \left(\frac{1}{m_{\rho_i}^{2}}+\frac{1}{m_{a_i}^{2}}\right)  \frac {f^2-f_{0K+1}^2}{f^2} 
\label{DeltaSnonlocal}
\end{equation}
that is the generalization of the result of Ref.\cite{4DCHM}.
According to this formula $\Delta S$ can have any value for appropriate choices of $f_{0K+1}^2$. However 
in this case the potential is quadratically divergent and the resonances not even partially unitarize the scattering
of Goldstone bosons unless $|f_{0K+1}^2|\ll f^2$.

Next we consider other non-minimal terms. 
The result is more involved than the previous one because all non minimal terms except $f_{0 K+1}$ enter in the mass spectrum in a non trivial way.
\\
To be concrete we include in our theory only two vector resonances ($\rho_{1,2}$) and one coset resonance ($a_1$). 
This model is described in detail in appendix \ref{C}.  Allowing for a logarithmic divergence in the potential there are two non-minimal interactions
$f_{02}$ and $f_{13}$. For $f_{13}=0$ one can write a simple analytical formula
\begin{equation}
\label{S12}
\Delta S=4\pi v^{2} \left(\frac{1}{m_{\rho_1}^{2}}+\frac{1}{m_{a_1}^{2}}+\frac{1}{m_{\rho_2}^{2}} -\frac{2 f_{02}^{2}}{f^{2}m_{a_1}^{2}} \right)  
\end{equation}
This shows that the contribution to the $S$-parameter can be reduced and can even become negative by properly choosing $f_{02}$. Differently from the result of Ref. \cite{4DCHM} obtained for a single level of resonances, in this case negative contribution to $\Delta S$ does not correspond necessarily to $m_{a_1} < m_{\rho_1}$.
\begin{figure}[t]
\centering
 {\includegraphics[width=0.48\columnwidth]{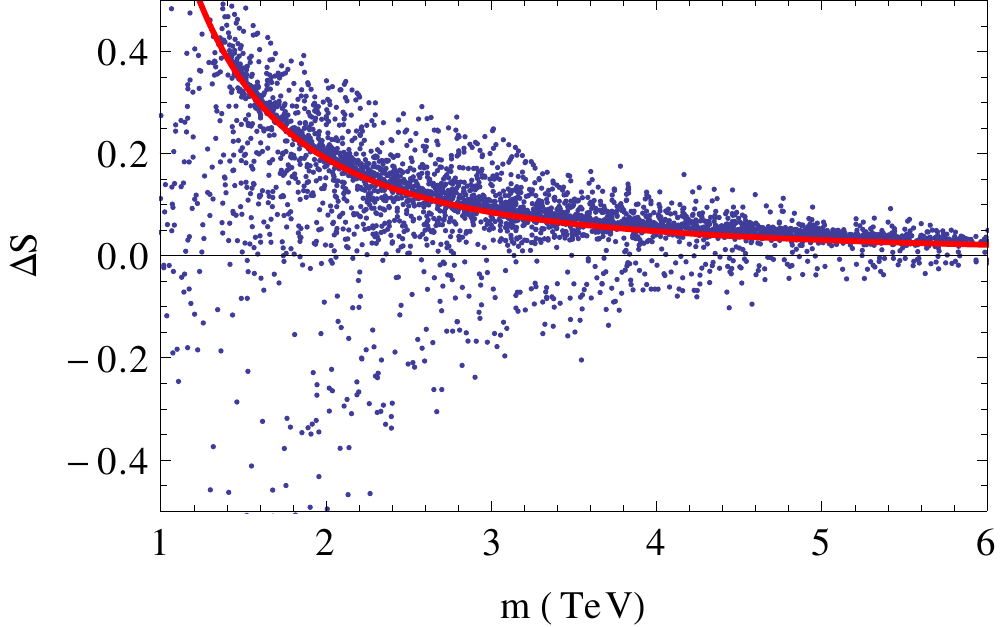}} \quad
 {\includegraphics[width=0.48\columnwidth]{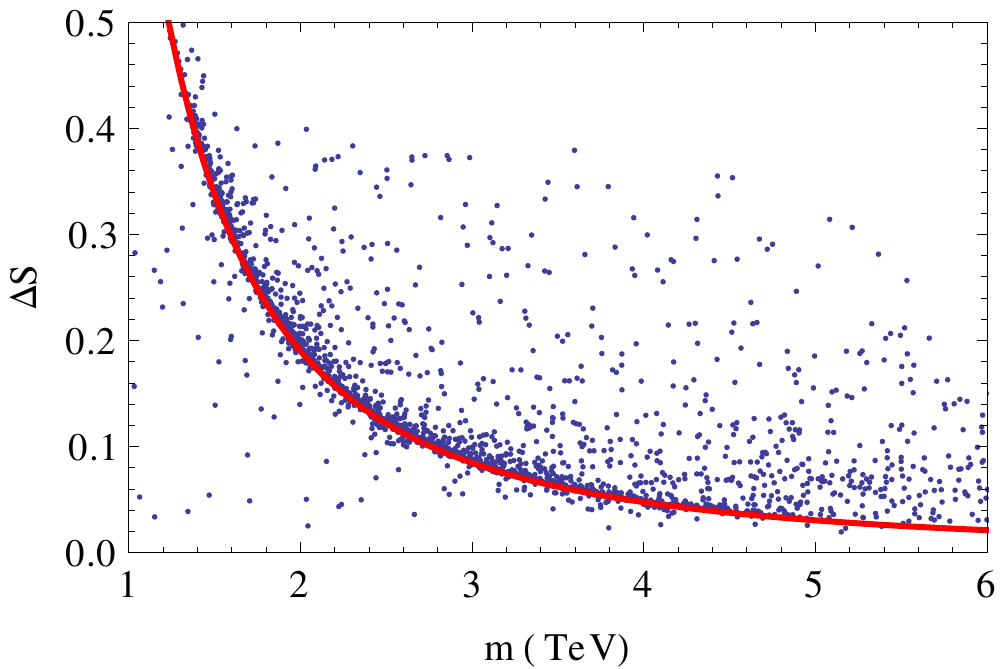}} \\
 \caption{\small{Tree level contribution from gauge resonances to  $S$  in the model with two vector resonances and a coset one with non minimal interaction $f_{02}$ (left) or $f_{13}$ (right).  The plot are obtained by scanning $f_{01}^2$, $f_{12}^2$ and $f_{02}^2$ or $f_{13}^2$ in the range $[-25,25] \, \text{TeV}^2$  and $g_1$ and $g_2$ in the range $[2,10]$ requiring that $f \simeq 1 \, \text{TeV}$. The curve $4 \pi v^2/m_*^2$ that minimizes $\Delta S$ with nearest-neighbour interactions is drawn in red. In the left plot lower and even negative values of $\Delta S$ are allowed and correspond to $f_{02}^2 > 0$. In the right plot $\Delta S$ can be smaller than in the nearest-neighbour case when $f_{13}^2 > 0$, but never negative.}}
 \label{fig:S-plot}
 \end{figure}
 
More in general we have performed a scan over the parameters of the model to determine the  allowed values for $\Delta S$ and plotted it as a function of the mass of the lightest resonance $m_*$. In the minimal case $\Delta S$ is minimized by the curve $4 \pi v^2/m_*^2$. Turning on the $f_{02}$ interaction, we get  lower and even negative contributions to $\Delta S$,  in agreement with the analytic formula \eqref{S12}.  
Performing  the same analysis for $f_{13}$, we get that small value of $\Delta S$ are allowed, but not negative ones. These results are reported in Fig.~\ref{fig:S-plot}. 
\\
We conclude that non minimal interactions can potentially lower the constraints on composite Higgs models coming from electroweak precision tests. We should mention however that this might come at the price of lowering the cut-off of the effective theory \cite{thaler}.

\section{Application to QCD}
\label{appQCD}

The formalism developed in this paper can be applied to describe in the most  general way 
the interaction of hadrons in QCD. The symmetry breaking pattern of QCD with two massless flavour is $SU(2)_L\times SU(2)_R/SU(2)_{L+R}$ with 
a discrete symmetry that exchanges $L \leftrightarrow R$.  This is locally equivalent to $SO(4)/SO(3)$ so that the formulas 
presented  in the previous sections also apply to this case.

We will here consider pions and vector mesons. Our approach is closely related 
to the one of AdS/QCD \cite{adsqcd1,adsqcd2}. We will follow and extend Ref.\cite{becciolini} where 4D models 
with nearest-neighbour interactions were studied, analyzing the physical consequences 
of non minimal terms. Related work can be found in \cite{wacker}.

\subsection{Electromagnetic splitting of pions and the KSRF relation}

Experimentally the mass difference between charged and neutral pions is
\begin{equation}
(m_{\pi^+}-m_{\pi^0})|_{\rm EXP}\simeq 4.6 \, {\rm MeV} \,.
\label{splittingexp}
\end{equation}
This difference is due to the explicit breaking of the global symmetry generated 
by the gauging of electromagnetism (the effect due to different quark masses being subleading).
The potential takes the form,
\begin{equation}
V(\pi^+ \pi^-)_{em} = \frac{3}{2}\int \frac{d^4Q}{(2\pi)^4}\log \left[1 + \frac{\Pi_1(Q^2)}{\Pi_0(Q^2)} \sin^2{\frac{\pi}{f_\pi}} \, \frac{\pi^+ \pi^-}{\pi^2} \right]
\end{equation}
where $\Pi_1$ and $\Pi_0$ are defined as in \eqref{Leff} and $f_\pi \simeq 92 \, \text{MeV}$ is the pion decay constant. To extract a quantitive prediction we have to resort to a specific model. 
A simple option is to consider the theory with one multiplet of gauge resonances with nearest-neighbour interactions, as in Ref.\cite{4DCHM}. 
This corresponds to an effective theory of QCD mesons, including only the lightest vector ($\rho$(770)) and axial ($a_1$(1260)) resonances. 
This is the minimal set of degrees of freedom that generates a finite potential. One finds,
\begin{equation}
m_{\pi^+}^2-m_{\pi^0}^2\simeq \frac {3 \alpha_{EM}}{4\pi} \frac {m_\rho^2 m_{a_1}^2}{m_{a_1}^2-m_\rho^2}\log\left( \frac {m_{a_1}^2}{m_\rho^2}\right)
\end{equation}
in agreement with Eq.~(\ref{splittingexp}) within about $25\%$.

A certain tension however exists between this result and other observables in low energy QCD.  One can parametrise,
\begin{equation}
\label{KSRF}
m_\rho^2= a \, g_{\rho\pi\pi}^2 f_\pi^2\,.
\end{equation}
where $g_{\rho\pi\pi}$ is the coefficient of the operator $\epsilon^{ijk} \rho_\mu^i \pi^j \partial^\mu \pi^k$.
Experimentally $a\simeq 2$ and the relation \eqref{KSRF} is known as the KSRF relation, see \cite{bandokugo}. 
In the model above that well reproduces  the electro-magnetic splitting of pions,  one finds 
\begin{equation}
\label{grhopipi_minimal}
g_{\rho \pi \pi}^2=\frac{m_\rho^2 (m_{a_1}^2 - m_\rho^2) (m_{a_1}^2 + m_\rho^2)^2}{4 f_\pi^2 m_{a_1}^6}
\end{equation}
corresponding to $a\simeq 3.4$ for the physical values of the masses (or 4 in the limit $m_{a_1}\to \infty$ \cite{georgi}).
Moreover the value $a\simeq 2$ cannot be reproduced with any number of resonances when only nearest-neighbour interactions are included. Indeed one can prove the following sum rule \cite{becciolini}
\begin{equation}
\label{rule}
f_\pi^2 \sum_{n=1}^K \frac{g_{\rho_n \pi\pi}^2}{m_{\rho_n}^2} = \frac{1}{3} - \frac{1}{12}\sum_{i=1}^{K+1} \frac{f_\pi^6}{f_{i-1i}^6} - \frac{1}{4} \frac{f_\pi^6}{f_{KK+1}^6} 
\end{equation}
implying $a>3$ for nearest-neighbour interactions.

This suggests that non minimal interactions should be relevant for hadrons. 
Adding the non-minimal term to the model above with a single $SO(4)$ multiplet Eq.~\eqref{grhopipi_minimal} is  modified into
\begin{equation}
g_{\rho \pi \pi}^2=\frac{(f_\pi^2-f_{02}^2) m_\rho^2 (m_{a_1}^2 - m_\rho^2) (m_{a_1}^2 + m_\rho^2)^2}{4 f_\pi^4 m_{a_1}^6}
\end{equation}
and the phenomenological value $a \simeq 2$ can be reproduced with $f_{02}^2 \simeq - 0.65 f_\pi^2$ ($f_{02}^2 = -f_{\pi}^2$ in the limit $m_{a_1}\to \infty$ considered in Ref. \cite{georgi}). 
As explained in section \ref{sec:potential} the non-minimal interactions generates in this case a quadratically divergent potential
\begin{equation}
m_{\pi^+}^2-m_{\pi^0}^2 = \frac {3 \alpha_{EM}}{4\pi} \frac{f_{02}^2}{f_\pi^2} \ \Lambda^2 \,.
\end{equation}
Physically we expect the integral to be cut-off by the heavier resonances. Since the next lightest vector resonance 
that could play a role is the $\rho'(1450)$ with mass roughly twice $m_\rho$, this spoils the agreement 
with (\ref{splittingexp}).

Our result is that with more resonances it is possible to simultaneously reproduce both the electro-magnetic splitting of pions 
and  the KSRF relation. We here present the simplest possibility where we include $\rho(770)$, $a_1(1260)$, $\rho'(1450)$. 
An effective description of these resonances is given by the model of appendix \ref{C}. We include two non-minimal terms $f_{02}$ and $f_{13}$
setting  $f_{03}$ to zero. The first term leads to a mild logarithmic divergence in the potential that we can assume to be cut-off by 
the heavier resonances. With a numerical analysis we find that (\ref{splittingexp}) and the KSRF relation can be reproduced 
for a reasonable choice of parameters. 
For example, the following choice of parameters roughly reproduces the phenomenology  
\begin{equation}
f_{01}^2 \sim 5 f_{\pi}^2, \quad f_{12}^2 \sim f_{\pi}^2, \quad f_{02}^2 \sim \frac{3}{2} f_{\pi}^2, \quad f_{13}^2 \sim -\frac{3}{2} f_{\pi}^2, \quad g_1 \sim 9, \quad g_2 \sim 8 \,,
\end{equation}
In fact,  with a cut-off $\Lambda\sim 2\,{\rm GeV}$, we get  $a \simeq 2$ and $m_{\pi^+}-m_{\pi^0}\simeq 4 \, {\rm MeV}$ and we also verify that  $g_{\rho \pi \pi} \sim g_\rho$, where $g_\rho$ 
is the trilinear self-coupling between the $\rho(770)$. This respects the coupling universality hypothesis of QCD but does not follow in general
for different choices of parameters in our lagrangian.  A detailed study of the other observables in low energy QCD will appear elsewhere.

\subsection{$L_9$ vs. $L_{10}$}
\label{sec:L9L10}

Let us finally  discuss the pion chiral lagrangian.
The effective lagrangian for  pions, up to fourth order in  derivatives is customarily
parametrised as follows\footnote{For the case $SU(2)\times SU(2)/SU(2)$ one combination of $L_1$, $L_2$ and $L_3$ is not independent \cite{pich1}.
We use this redundant notation as the final formula holds in general for $SU(N)\times SU(N)/SU(N)$.} 
\begin{eqnarray}
 \mathcal{L}_{p^2} &=& \frac{f_\pi^2}{4} \rm{Tr}[\hc{(D_\mu \Sigma)} (D^\mu \Sigma)]\,, \\
 \mathcal{L}_{p^4} &=&
 L_1 \rm{Tr}[\hc{(D_\mu \Sigma)} (D^\mu \Sigma)]^2
 + L_2 \rm{Tr}[\hc{(D_\mu \Sigma)} (D_\nu \Sigma)] \rm{Tr}[\hc{(D^\mu \Sigma)} (D^\nu \Sigma)] \\
 && +\ L_3 \rm{Tr}[\hc{(D_\mu \Sigma)} (D^\mu \Sigma) \hc{(D_\nu \Sigma)} (D^\nu \Sigma)] \notag\\
 && -\ \ii L_9 \rm{Tr}[\mathit{l}_{\mu\nu}(D^\mu\Sigma)\hc{(D^\nu\Sigma)} +
 \mathit{r}_{\mu\nu}\hc{(D^\mu\Sigma)}(D^\nu\Sigma) ]
 + L_{10} \rm{Tr}[\hc{\Sigma}\mathit{l}_{\mu\nu}\Sigma\mathit{r}^{\mu\nu}]\,, \notag
\label{chiralL}
\end{eqnarray}
where $\Sigma$ is a unitary matrix and the covariant derivatives with respect the sources $l_\mu$ and $r_\mu$ 
are defined as
\begin{equation}
D_\mu \Sigma \equiv \partial_\mu \Sigma - i l_\mu \Sigma+ i \Sigma r_\mu \,.
\end{equation}

The coefficients obtained by integrating out the resonances can be determined as in section \ref{GBlagrangian}.
With the leading non-minimal interactions one finds\footnote{The coefficients can be extracted from eqs. (\ref{so5coefficients})
using the identities in the appendix of \cite{continovector}. One finds  $L_1=- c_1/2\,, L_2=c_2 \,, L_3=-3 c_2\,, L_9 = - c_4^+\,, L_{10}= -2 c_5^+$.},
\begin{equation}
\begin{split}
&L_1=\frac 1 2 L_2=-\frac 1 6 L_3= \sum_{i=1}^K \frac{[1-\alpha_i^2]^2}{16 g_i^2} \\
&L_9 = -L_{10}=\sum_{i=1}^K \frac{[1-\alpha_i^2]}{2 g_i^2} \,.
\end{split}
\end{equation}
where $\alpha_i$ are defined as in Eq.~\eqref{alpha_i}.
This implies the relations
\begin{equation}
\begin{split}
2 L_1-L_2=0\,,&~~~~~~~~~~~~~~~~~3L_2+L_3=0\\
&L_9 + L_{10}=0
\label{L9L10}
\end{split}
\end{equation}
valid at tree level. Experimentally \cite{pich1}:
\begin{equation}
\frac {L_9 + L_{10}}{L_9-L_{10}}= 0.1 \pm 0.1
\end{equation} 
in good agreement with Eq.~(\ref{L9L10}). The other relations in (\ref{L9L10}) are also satisfied with similar accuracy.
$L_9+L_{10}\simeq 0$ is spoiled by non minimal kinetic terms in our description.\footnote{Higher derivatives terms such as  ${\rm Tr}[\rho^{\mu\nu} i [d_\mu,d_\nu ]]$  would also modify this relation, see \cite{continovector,Azatov:2013ura}.} In this case the tree level contribution from the exchange of
axial resonances violates the last relation in Eq.~(\ref{L9L10}), see Eq.~(\ref{so5coefficientsnew}).
If the resonances are weakly coupled one could however expect such terms to be suppressed 
for the consistency of the effective theory.

This generalises the result found in theories with nearest-neighbour interactions \cite{pw5D,becciolini}. 
For large $N$ theories with weakly coupled 5D duals one can prove that the corrections to the relations 
above are small and in fact vanish in the large $N$ limit. In general confining gauge theories it is not a priori 
clear that spin-1 resonances should be described as gauge fields and therefore $L_9+L_{10}\simeq 0 $ 
may not follow \cite{pich2}. 

\section{Conclusions}
\label{conclusions}

In this paper we have presented a new parametrisation of composite resonances in theories with spontaneously broken 
global symmetries. The construction generalizes nearest-neighbour interactions reminiscent of extra-dimensional theories
to reconstruct the most general lagrangian compatible with the symmetries. Our approach allows to
systematically characterize the deviations from extra-dimensional theories. 
This can be encoded into the notion of locality in theory space: nearest-neighbour interactions maximize the distance 
between elementary fields and the dynamics that breaks spontaneously the symmetry while non-minimal terms 
shorten this distance.

The physical motivation for this work was two-fold. On one hand we wanted to explore the most general possibilities 
allowed by strong dynamics in the context of composite Higgs models where the Higgs is a Goldstone boson.
These models have been mostly studied  in the context of 5D realizations or in the extreme limit where only 
one multiplet of resonances is light and one considers the most general effective lagrangian compatible with the symmetries.
As we have seen the presence of non-minimal terms allows to interpolate between these descriptions.
Indeed we recover as special limits various effective descriptions considered in the literature.
We also show that non-minimal terms have important physical consequences affecting the calculability of the GB potential 
and the UV behaviour of the theory.  They also allow to deviate from results in extra-dimensional theories where for example the tree level contribution to the 
$S$-parameter is always positive and typically large. In our more general setup any small or even negative values of $\Delta S$ 
could be reproduced.

Secondly our approach is suitable for a general parametrisation of hadrons. In the last ten years it has been shown 
that extra-dimensional theories approximate reasonably well several low energy QCD data \cite{adsqcd1,adsqcd2}. 
This is remarkable because QCD is not a conformal field theory where the AdS/CFT correspondence can be applied.
Some observables however are not reproduced with good accuracy. For example the KSRF relation appears in tension with 
the nearest-neighbour interactions hypothesis. We have shown that non-minimal terms are relevant in this regard
to reproduce the KSRF relation compatibly with the electro-magnetic splitting of pions in QCD.
We have also shown that the experimental relation $L_9\approx - L_{10}$ among the parameters of the chiral lagrangian 
follows in general if the resonances are treated as gauge resonances with the leading interactions.  
We hope to return to a systematic  study of these and  related questions in the near future.

\vspace{1cm}

\subsection*{Acknowledgments}
The work of MR is supported in part by the MIUR-FIRB grant RBFR12H1MW. We would like to thank 
Roberto Contino, Gero Von Gersdorff, Giuliano Panico, Alex Pomarol, Eduardo Ponton, Rogerio Rosenfeld for interesting discussions
and especially Andrea Tesi for collaboration at the early stages of this work.

\newpage

\appendix

\section{Fermionic Form Factors} 
\label{B}

In this appendix we collect the relevant formulas for the fermionic sector, following closely \cite{4DCHM} to which we refer for details.
The effective lagrangian for composite fermions in the $\mathbf{5}$ rep of $SO(5)$ ($\text{CHM}_{\mathbf{5}}$) in a constant GB background, 
takes the form
\begin{eqnarray}
\label{CHM5-effA}
{\cal L}^{{\rm CHM_5}}_{{\rm eff}} &=&\ \!
 \bar{\Psi}_{0_L}^\alpha \slashed{p} \left( \delta^{\alpha \beta} \hat{\Pi}^{q_L}_0(p^2)
  + \Phi^\alpha \Phi^\beta \hat{\Pi}_1^{q_L}(p^2) \right)\! \Psi_{0_L}^\beta
  + \! \bar{\Psi}_{0_R}^\alpha \slashed{p} \left( \delta^{\alpha \beta}\hat{\Pi}_0^{u_R}(p^2)
  + \Phi^\alpha \Phi^\beta \hat\Pi_1^{u_R}(p^2) \right)\! \Psi_{0_R}^\beta \nonumber \\
 &+& \bar{\Psi}_{0_L}^\alpha \!\left( \delta^{\alpha \beta} \hat{M}_0^{u}(p^2)
   +\Phi^\alpha \Phi^\beta \hat{M}_1^{u}(p^2) \right)\! \Psi_{0_R}^\beta
 + h.c.
\end{eqnarray}
where $\Psi^0_{L,R}$ are the sources of fermionic operators in the $\mathbf{5}$ rep.
The form factors can be written in terms of the correlation functions of the fermionic fields. 
The fundamental representation of $SO(5)$ decompose under $SO(4)$ in the fundamental and a singlet, i.e. $\mathbf{5}=\mathbf{4} \oplus \mathbf{1}$. 
The sources $\Psi^0_{L,R}$ decompose in fourplets and singlets, so we can define six correlation functions: $\Pi_{LL,RR}^{\mathbf 4}$, $\Pi_{LL,RR}^{\mathbf 1}$ and $\Pi_{LR}^{\mathbf 4}$, $\Pi_{LR}^{\mathbf 1}$. 
The relations between the form factors that appear in \eqref{CHM5-effA} and the fermionic 
correlation functions are
\begin{align}
\label{fermionic_ff}
&\hat{\Pi}_{0}^{q_L}=\Pi_{LL}^{\mathbf 4} \,, \quad \quad 
\hat{\Pi}_{0}^{u_R}=\Pi_{RR}^{\mathbf 4} \notag \\
&\hat{\Pi}_{1}^{q_L}=\Pi_{LL}^{\mathbf 1}-\Pi_{LL}^{\mathbf 4} \,,  \quad  \quad 
\hat{\Pi}_{1}^{u_R}=\Pi_{RR}^{\mathbf 1}-\Pi_{RR}^{\mathbf 4} \notag \\
&\hat{M}_{0}^{u}=\Pi_{LR}^{\mathbf 4} \,, \quad \quad \hat{M}_{1}^{u}=\Pi_{LR}^{\mathbf 1}-\Pi_{LR}^{\mathbf 4}
\end{align}
For simplicity we consider the case when the sources couple to the nearest resonance.
Integrating out the composite fermions, we obtain
\begin{align}
\label{fermionic_corr}
&\Pi_{LL}^{\mathbf{4}}(p^2)=-|M_{01}|^{2}\left[(p^{2}I-{\cal M}_{Q}^{\dag}{\cal M}_{Q})^{-1}\right]_{11}   \notag \\
&\Pi_{RR}^{\mathbf{4}}(p^2)=-|M_{10}|^{2}\left[(p^{2}I-{\cal M}_{Q}{\cal M}_{Q}^{\dag})^{-1}\right]_{11}  \notag \\
&\Pi_{LR}^{\mathbf{4}}(p^2)=-M_{01}M_{10} \left[(p^{2}I-{\cal M}_{Q}^{\dag}{\cal M}_{Q})^{-1}{\cal M}_{Q}^{\dag}\right]_{11} 
\end{align}
while the correlators $\Pi_{LL}^{\mathbf{1}}$, $\Pi_{RR}^{\mathbf{1}}$, $\Pi_{LR}^{\mathbf{1}}$ can be found by replacing ${\cal M}_{Q}\rightarrow {\cal M}_{S}$, $M_{01}\rightarrow M_{01} + Y_{01}$ and $M_{10}\rightarrow M_{10} + Y_{10}$.

Finally, we can write explicitly the effective lagrangian that describes the coupling of the SM fermions to the Higgs: 
\begin{equation}
\label{Leff-fermions}
\begin{split}
{\cal L}&=\bar{q}_{L}~\slashed p~\left( \Pi_{0}^{q}(p^{2})+\frac{1}{2} s_{h}^{2} \Pi_{1}^{q}(p^{2}) \hat{H}_{c} \hat{H}_{c}^{\dag} \right) q_{L} + \bar{t}_{R}~ \slashed p~\left( \Pi_{0}^{t}(p^{2})+\frac{1}{2} s_{h}^{2} \Pi_{1}^{t}(p^{2})\right) t_{R} \\
&+\frac{s_{h}c_{h}}{\sqrt{2}}  M_{1}^{t}(p^{2}) ~\bar{q}_{L} \hat{H}_{c} t_{R} + \text{h.c.}
\end{split}
\end{equation}
where $s_h=\sin{h/f}$ and $c_h=\cos{h/f}$.
We recall also that $\hat{H}^{c}=i\sigma_{2}\hat{H}^{*}$ and \mbox{$\hat H=1/h(h^2+ih^1, h^4-i h^3)^T$}. The form factors that appear in the effective Lagrangian \eqref{Leff-fermions} 
are related to those of Lagrangian \eqref{CHM5-effA} by
\begin{align}
\label{fermionic_ff_kin}
&\Pi_{0}^{q}=\frac{1}{y_{tL}^{2}} + \hat{\Pi}_{0}^{q_L}  \quad , \quad 
\Pi_{0}^{t}=\frac{1}{y_{tR}^{2}} + \hat{\Pi}_{0}^{u_R} + \hat{\Pi}_{1}^{u_R}   \notag \\
&\Pi_{1}^{q}=\hat{\Pi}_{1}^{q_L} \quad , \quad 
\Pi_{1}^{t}=-2 \hat{\Pi}_{1}^{u_R}   \quad , \quad 
M_{1}^{t}= \hat{M}_{1}^{u} 
\end{align}
where $y_{t_{L,R}}$ come from adding the kinetic terms for the elementary fermions. 
The fermionic contribution to the Higgs potential, derived from the Lagrangian \eqref{Leff-fermions}, is $(Q^2=-p^2)$:
\begin{equation}
\label{fermion pontential}
V(h)_{top}=-2 N_c \int \frac{d^4Q}{(2\pi)^4}\log \left[\left(1 + \frac {\Pi_1^q}{2 \Pi_0^q} s_h^2\right)\left(1 + \frac {\Pi_1^t}{2 \Pi_0^t} s_h^2\right)+
\frac {|M_1^t|^2}{2 Q^2 \Pi_0^q \Pi_0^t} s_h^2 c_h^2 \right]  \,.
\end{equation} 
The form factors $\Pi_0^q$ and $\Pi_0^t$ are dominated in the UV by the kinetic terms for the elementary fermion fields, see Eq.~\eqref{fermionic_ff_kin}. 
Using this fact, one derives the condition (\ref{weinbergsumrules}) for the convergence of the potential.

\section{An Explicit Example}
\label{C}

In this appendix we collect some explicit formulas valid for the model \eqref{Lag_NM} with $K=2$ and $f_{23} \rightarrow\infty$. 
The model with $K=2$ describes two complete multiplets of $SO(N)$ resonances. In the $f_{23} \rightarrow\infty$ limit,
the last coset resonance decouples and the model is an effective description of two resonances in the adjoint of $SO(N-1)$ 
and a coset resonance transforming as the fundamental of $SO(N-1)$. Neglecting non minimal kinetic terms, 
the Lagrangian can be easily written  in the CCWZ gauge starting from \eqref{LagCCWZ}:
\begin{equation}
\begin{split}
&{\cal L} = \frac{f_{01}^2}{4} \text{Tr}[(e^\mu - \rho_1^\mu)^2]+\frac{f_{02}^2}{4} \text{Tr}[(e^\mu - \rho_2^\mu)^2]+ \frac{f_{01}^2}{4} \text{Tr}[(d^\mu - a_1^\mu)^2]+\frac{f_{12}^2}{4} \text{Tr}[(\rho_{1}^\mu - \rho_{2}^\mu)^2] \\
&+ \frac{f_{12}^2+f_{13}^2}{4} \text{Tr}[(a_{1}^\mu)^2] + \frac{f_{02}^2+f_{03}^2}{4} \text{Tr}[(d^\mu)^2] 
-\frac{1}{4 g_{1}^2} \text{Tr} \left[(\partial_\mu \rho_{1\nu}-\partial_\nu \rho_{1\mu}-i[\rho_{1\mu},\rho_{1\nu}]-i[a_{1\mu},a_{1\nu}])^2 \right] \\
&-\frac{1}{4 g_{2}^2} \text{Tr} \left[(\partial_\mu \rho_{2\nu}-\partial_\nu \rho_{2\mu}-i[\rho_{2\mu},\rho_{2\nu}])^2 \right] 
-\frac{1}{4 g_{1}^2} \text{Tr} \left[(\partial_\mu a_{1\nu}-\partial_\nu a_{1\mu}-i[\rho_{1\mu},a_{1\nu}]-i[a_{1\mu},\rho_{1\nu}])^2 \right] \,.
\end{split}
\end{equation}
Integrating out the resonances, we obtain the form factors
\begin{align}
&\Pi_0(p^2)= -\frac{p^2}{g_0^2} + \frac{p^2 \left(2 p^2 \left(f_{01}^2+f_{02}^2\right)-\left(g_1^2+g_2^2\right) \left(f_{01}^2
   \left(f_{02}^2+f_{12}^2\right)+f_{02}^2 f_{12}^2\right)\right)}{-2 p^2 \left(g_1^2
   \left(f_{01}^2+f_{12}^2\right)+g_2^2 \left(f_{02}^2+f_{12}^2\right)\right)+g_1^2 g_2^2 \left(f_{01}^2
   \left(f_{02}^2+f_{12}^2\right)+f_{02}^2 f_{12}^2\right)+4 p^4} \\
\begin{split}
&\Pi_1(p^2)=-\frac{2 p^2 \left(2 p^2 \left(f_{01}^2+f_{02}^2\right)-\left(g_1^2+g_2^2\right) \left(f_{01}^2
   \left(f_{02}^2+f_{12}^2\right)+f_{02}^2 f_{12}^2\right)\right)}{-2 p^2 \left(g_1^2
   \left(f_{01}^2+f_{12}^2\right)+g_2^2 \left(f_{02}^2+f_{12}^2\right)\right)+g_1^2 g_2^2 \left(f_{01}^2
   \left(f_{02}^2+f_{12}^2\right)+f_{02}^2 f_{12}^2\right)+4 p^4} \\
&-\frac{f_{01}^4 g_1^2}{g_1^2
   \left(f_{01}^2+f_{12}^2+f_{13}^2\right)-2 p^2}+f_{01}^2+f_{02}^2+f_{03}^2 \,.
\end{split}
\end{align}
The GB decay constant is
\begin{equation}
f^2=\Pi_1(0)=f_{01}^2+f_{02}^2+f_{03}^2-\frac{f_{01}^4}{f_{01}^2+f_{12}^2+f_{13}^2}
\end{equation}
while the poles of $\Pi_1(p^2)$ give the masses of the physical resonances
\begin{align}
&m_{\rho_{1,2}}^2 = \frac{1}{4} \Bigl(g_1^2 f_{01}^2  + g_2^2 f_{02}^2 +g_1^2 f_{12}^2
   +g_2^2 f_{12}^2  \\
   &\mp\sqrt{g_1^4 f_{01}^4 +2 g_1^2  f_{01}^2  \left(\left(g_1^2-g_2^2\right) f_{12}^2
   - g_2^2 f_{02}^2 \right) + g_{2}^4 f_{02}^4 +2 g_{2}^2
   \left(g_{2}^2-g_{1}^2\right) f_{02}^2 f_{12}^2  
   +\left(g_{1}^2+g_{2}^2\right)^2 f_{12}^4 } \Bigr) \notag \\
&m_{a _1}^2 = g_1^2 \left(\frac{f_{01}^2}{2} + \frac{f_{12}^2}{2}+\frac{f_{13}^2}{2}\right) \,.
\end{align} 
Let us now consider the integral 
\begin{equation}
\int^\Lambda d^4 Q \ \frac{\Pi_1(Q^2)}{\Pi_0(Q^2)}
\end{equation}
that controls the GB potential due to gauge loops.
It is easy to verify that the terms $f_{03}$ and $f_{02}$ cause  a quadratic and logarithmic dependence on the cut-off, respectively. This can be shown by following the argument of section \ref{sec:potential}. Consider the model with $f_{23}$ finite. The link $f_{03}$ identifies a path with vanishing length corresponding to the quadratic dependence on the cut-off . 
The links $f_{02}$ and $f_{13}$ both identify a path of unit length leading to a cut-off independent result. 
Since the path defined by $f_{02}$ includes the last link, in the limit $f_{23} \rightarrow \infty$ it leads to a logarithmic dependence on the cut-off.

\bibliography{arc}{}
\bibliographystyle{utphys}

\end{document}